\documentclass[12pt]{iopart}

\usepackage{graphicx}

\begin{document}

\title[ Solar cell wavelength response]{Investigation of low band gap silicon alloy thin film solar cell for improving short and long wavelength response}

\author{S. M. Iftiquar, J. Yi}

\address{College of information and Communications Engineering, Sunkyunkwan University, Suwon, South Korea }
\ead{smiftiquar@gmail.com}

\begin{abstract}

Numerical simulation of a solar cell can provide various information that can be useful to maximize its power conversion efficiency (PCE). In that respect we carried out a set of numerical simulation using AFORS-HET simulation program. Separately, in order to get a better understanding, the optical absorption in individual layers devices were analyzed. Current-voltage characteristic curve of a reference cell (Cell-A) was used as the starting device. The PCE of the reference device was $8.85\%$ with short circuit current density $J_{sc}$ of 15.43 mA/cm$^{2}$ and fill factor (FF) of $68.3\%$. However, it was noticed that the reference cell had high parasitic optical absorption at the window layer and the device structure was also not optimized. After suitable optimization the PCE of this device (Cell-B2) improves to $11.59\%$ ($J_{sc}$ and FF of 13.0 mA/cm$^{2}$ and $87\%$ respectively). The results show that the effective optical absorption in the active layer can be improved significantly by optimizing the device structure. The short wavelength response can be improved by reducing the parasitic optical absorption by the doped window layer, while its long wavelength response improves by raising effective absorption length of the active layer. Furthermore, its optimum thickness, for the highest possible PCE, is found to be dependent upon the material properties, more importantly on its defect density. 

\vspace{2pc}
\noindent{\bf keywords}: Thin film; amorphous silicon; solar cell; light absorption; power conversion efficiency  

\end{abstract}
 
\maketitle

\section{Introduction}

Photovoltaic conversion efficiency (PCE) of an amorphous silicon thin film solar cell is often limited by carrier recombination in mid-gap defects of the active layer. While improved film deposition technique such as hydrogen dilution can be used to reduce the defect density to about $10^{15}$ cm$^{-3}$ and get the PCE close to $10\%$, yet the ability to reduce the defect density by hydrogen dilution appears as a fundamental limit to the improvement of material quality. In addition, when high hydrogen dilution is used its flow rate determines a reduced film deposition rate, leading to a longer time to fabricate a device. 

\subsection{Silicon oxide window layer}

Additionally, the doped window layer contributes to parasitic optical absorption, which is a loss of light thereby the PCE get further limited. While strategies of applying a wide band gap silicon oxide (a-SiO:H) layer can in principle raise the PCE yet the reported PCE also appear to have limit.

 Furthermore, the work function of a wider band gap a-SiO:H layer may lead to band discontinuity, for which a suitable double layer is also suggested, however, this second layer contributes to an increased parasitic optical absorption. It is therefore desirable to develop strategies to improve device performance even with a moderate defect density of the active layer and thinner window layer.
 
 Moderate defect density can be easily maintained  with an existing plasma enhanced chemical vapor deposition (PECVD) system, but using a thinner
 window layer requires an improved PECVD system with less surface roughness or a high degree of  uniformity of the deposited film. For example, if a maximum surface roughness of a film is 7 nm, then  depositing film of thickness 7 nm or less will result in a faulty device. However, a state of the art PECVD  system can be used to fabricate devices with film thickness 7 nm or less. In order to analyze importance  of such thin layers we carried out the investigation

\subsection{Simulation}

In this work, we demonstrate that a high device efficiency is achievable by using a moderate quality amorphous silicon layers. This device structure has been adapted from overall development of single junction and tandem solar cell by using wide band gap silicon alloy and the results are obtained by freely available AFORS-HET simulation program. More specifically, we use the Beer-Lambert absorption law and Fresnel's reflection to obtain incoherent optical absorption and transmittance of individual layers in a device structure. Here the absorbance is a key parameter and is generally known as the product of absorption coefficient and film thickness. It is generally known that a thicker film will absorb more light, therefore high current density can be expected with a thicker active layer. This can be true if the carrier transport across the film do not degrade significantly. But for amorphous silicon based device, the degradation is prominent due to the Schockley-Read-Hall type carrier recombination at the mid-gap defects. 

\subsection{Parasitic optical absorption}

Although there have been previous observations that thin amorphous silicon film are useful for solar cell device structure, the fact that the device performance can actually be improved with an ultrathin amorphous silicon film has not been well appreciated, as a large number of the reported devices contain thicker doped window layer ($> 10$ nm) and thicker active layer ($\approx 300$nm), where the device performance remains lower than that can be achieved with an optimized one. By contrast we will see that PCE of our optimized device is $> 10\%$ even with a relatively more defective active layer. The effect on its performance can be more prominent in a tandem device.  Comparing this with the other high efficiency thin film devices made with thin film perovskite material, where the device efficiency can be as high as $20\%$ or more, are not limited by parasitic optical absorption of the electron or hole collecting layers or defect density of the absorber layer. The micro or nano-crystalline perovskite material exhibits high carrier mobility, lower defect density, which is significantly lower in amorphous silicon based material. 

\subsection{Perovskite solar cell}

In contrast to the amorphous silicon based technology, the perovskite solar cell require (i) spin coating of organic-inorganic precursor material, (ii) thicker electron and hole transporting layer, (iii) small size of the fabricated device. It is to be noted that the long term stability of these devices is an well known issue and dimension of these devices are relatively small because of significant non-uniformity of spin-coated layers. These devices also exhibit a phenomena known as hysteresis, which is the difference in current density-voltage (J-V) characteristic curves in the forward and reverse voltage sweep. It was also reported that at a different speed of the voltage sweep the estimated PCE also differs. Under AM1.5 illumination, the perovskite device absorbs shorter wavelength radiation as its optical absorption band is below 700 nm wavelength. Therefore, one of the most popular application of the perovskite solar cell is tandem structure, with a crystalline silicon (c-Si) heterostructure (HIT) device at the bottom sub-cell. Under the AM1.5 illumination, it becomes easier to match the current density of the top and bottom sub-cells to a higher value as the perovskite solar cell can generate a very high current density. Since the perovskite device is composed of different material than the silicon based alloy, the interface formation between the top perovskite and the bottom HIT sub-cells remains a challenge. On the other hand earlier research work has shown that the silicon alloy based single and tandem devices can exhibit high PCE. 

\subsection{Various silicon alloy}

Extensive research has been going on to improve performance of photovoltaic solar cells. It includes the development of a large number of device structures as well as developing various  materials. Various silicon alloy materials are found to be useful in thin film silicon solar cell, of which intrinsic type hydrogenated amorphous silicon germanium (i-a-SiGe:H) [3-5] and boron doped p-type amorphous silicon oxide (p-a-SiO:H) or p-type nano-crystalline silicon oxide (p-nc-SiO:H) [2, 6, 7], p-type amorphous silicon (p-a-Si:H), p-type nano-crystalline silicon (p-nc-Si:H) are few examples. One of the primary purposes of investigating solar cell is to achieve higher power conversion efficiency (PCE), which is generally measured under AM1.5G insolation. Higher the PCE of a device, higher is its capability to generate electrical energy from light.

In order to raise the PCE, the material of the layers and its device structure requires optimization. Optical properties of the layers determine amount of light absorbed while the electronic properties determine collection efficiency of photo generated charge carriers. It is generally believed that adding oxygen to the amorphous silicon network improves its optical band gap or optical transparency [8-10], therefore the silicon oxide based doped window layer is expected to improve light transmission into the absorber layer, hence increase current density. For that purpose p-type silicon oxide layer is used and optimized in various solar cells. 

Although i-a-SiGe:H has been investigated for a very long time [3-5] yet a further improvement in its optoelectronic properties appears to have reached a saturation.   The published literature suggests that the i-a-SiGe:H is more defective material [3-5], because of which the reported performance of a device using this photosensitive has been low [11-13]. Defects in the i-a-SiGe:H may have insignificant effect on device performance [14]. Because of higher optical absorption in the i-a-SiGe:H material [3-5, 15], its use in a solar cell is expected to improve the long wavelength response of a device.

Various other approaches have also been adopted to achieve a higher PCE, like high quality material with low defect density, high optical absorption in the active or absorber region while reducing optical absorption loss at the doped window layers. Optical absorption in the doped layer can be reduced by using wide band gap materials like silicon oxide [2, 6, 7], silicon carbide. Furthermore, intrinsic hydrogenated amorphous silicon (i-a-Si:H) , hydrogenated amorphous silicon oxide (i-a-SiO:H) [8, 9] have also been investigated, among which the i-a-Si:H shows better characteristics for its application as active layer because of its lower defect density with reference to a-SiGe:H [3-5] or a-SiO:H [9]. In principle, lower optical gap of the active layer is useful for absorbing long wavelength radiation while the wide band gap material can be useful for absorbing short wavelength ones. However, it may be possible to extend the range of absorption wavelength in the active layer by changing design of the device as well. Here we report such an investigation. We used an i-a-SiGe:H based solar cell, estimated its current density-voltage characteristics (J-V curve) by numerical simulation. 

Silicon based p-i-n type solar cells have been one of the most popular devices [1, 2]. Here p-i-n indicates i-type active layer sandwiched between p-type and n-type doped layer. Wide band gap p-type window layer can enhance device performance partly because of a reduction in parasitic optical absorption. Efficiently doped layers can provide higher built-in field to separate the photo generated charge carriers. The charge carriers are generated in the active layer. Defects in the active layer can degrade device performance. Defect mediated Schockley-Read-Hall (SRH) recombination leads to reduced short circuit current density ($J_{sc}$), open circuit voltage ($V_{oc}$) and fill factor. As a result, defective active layer leads to a poor device performance. One of the primary reasons of low power conversion efficiency of amorphous silicon based solar cell is the high defect density of the active layer. 

This theoretical investigation was carried out with a reasonably acceptable material properties. Here, the simulation was started with a reference device structure (Cell-A). After that the device structure was altered and its resulting performance was investigated. 

The importance of this work is that it can give a clear indication about few of the important issues related to device performance.

 \section{Simulation}
 
Initial reference device structure used in the simulation is (Cell-A) as the following: Glass/ITO/p-nc-SiO:H/p-a-SiO:H/i-a-SiGe:H/a-Si:H/n-nc-Si:H/Ag, as shown in Fig. 1(a). Then the device structure was changed to Cell-B, by removing the p-nc-SiO:H layer. Here the i-a-Si:H is the intrinsic hydrogenated amorphous silicon layer, used as buffer layer to match the electronic band structure between the i-a-SiGe:H and the n-type nano-crystalline silicon (n-nc-Si:H) layers. Opto-electronic properties of the layers adapted for the simulation is shown in Table I. These values are adapted from various publications, although these values may not match exactly to experimentally measured values reported in the literatures, but it falls under the range of reported values. 

Optical absorption transmission of each of the layers are determined by its absorption coefficient $(\alpha)$, following Beer-Lambert's exponential relation. Light transmitted to the a-SiGe:H active layer ($I_{t}$) can be expressed as:

\begin{equation}
	I_{t}=I_{0} \prod_{(x=1)}^{5}(1-R_{x} )\exp(-\alpha_{x} d_{x})
\end{equation}

Here the symbol $\prod_{x=1}^{5}$ stands for product, with range of values for x from 1 to 5, with x=1 for air, x=2 for glass, x=3 for ITO, x=4 for p-nc-SiO:H, x=5 for p-a-SiO:H. Here $I_{0}$ is intensity of incident light, $R_{x}$ is Fresnel reflectivity at the interface, $d_{x}$ is film thickness while $\alpha_{ }$ is optical absorption coefficient of respective layers. The absorbed light ($I_{a}$) at the i-a-SiGe:H layer can be expressed as:

\begin{equation}
	I_{a}=I_{t} (1-\exp(-\alpha d))
\end{equation}

Here $\alpha$ is optical absorption coefficient and $d$ is thickness  of the i-a-SiGe:H layer. The above formulation can be adapted to estimate light absorption in other layers as well. 

The numerical simulation for the electronic characteristics of the device is estimated by using freely available AFORS-HET v2.4.1 simulation program [16]. The schematic structure of the device is shown in Fig. 1(a).   

In the AFORS-HET the operation of a solar cell is described by one dimensional semiconductor equations. These semiconductor equations are Poisson’s equation and transport equation for electrons and holes. These are also known as drift-diffusion equation for the electrons and holes in a semiconductor. In the AFORS-HET these equations are solved with the help of finite difference method under different conditions. At the first stage of the simulation, the equilibrium carrier densities were estimated under dark condition and without external bias. Then electron-hole pairs are generated due to absorption of incident photons. The absorption is described by Lambert-Beer expression by taking incoherent internal passage of transmitted light. Here Shockley-Read-Hall recombination were considered and sub-bandgap recombination was treated as the primary mechanism of loss of photo generated carriers. The interface currents are modeled by drift diffusion and here it is not changed by considering zero surface charge density. The contacts between the doped layers and transparent conducting oxides at the front and back of the device were modeled as Schottky metal-
semiconductor (MS Schottky) contacts.

The J-V characteristic curve of the Cell-A, as obtained from the simulation is as shown in Fig. 1(b).It’s open circuit voltage ($V_{oc}$ ) was 0.84 V, short circuit current density ($J_{sc}$ ) was 15.43 mA/cm$^{2}$ , fill factor (FF) is $68.25\%$ and PCE as $8.85\%$, which is approximately close to that reported in [16] as 0.86 V, 15.54 mA/cm$^{2}$ , $68.34\%$ and $9.13\%$ respectively (cell-D of Table 2 of the reference [16])

\begin{figure}
	\centering
	\includegraphics[width=6cm]{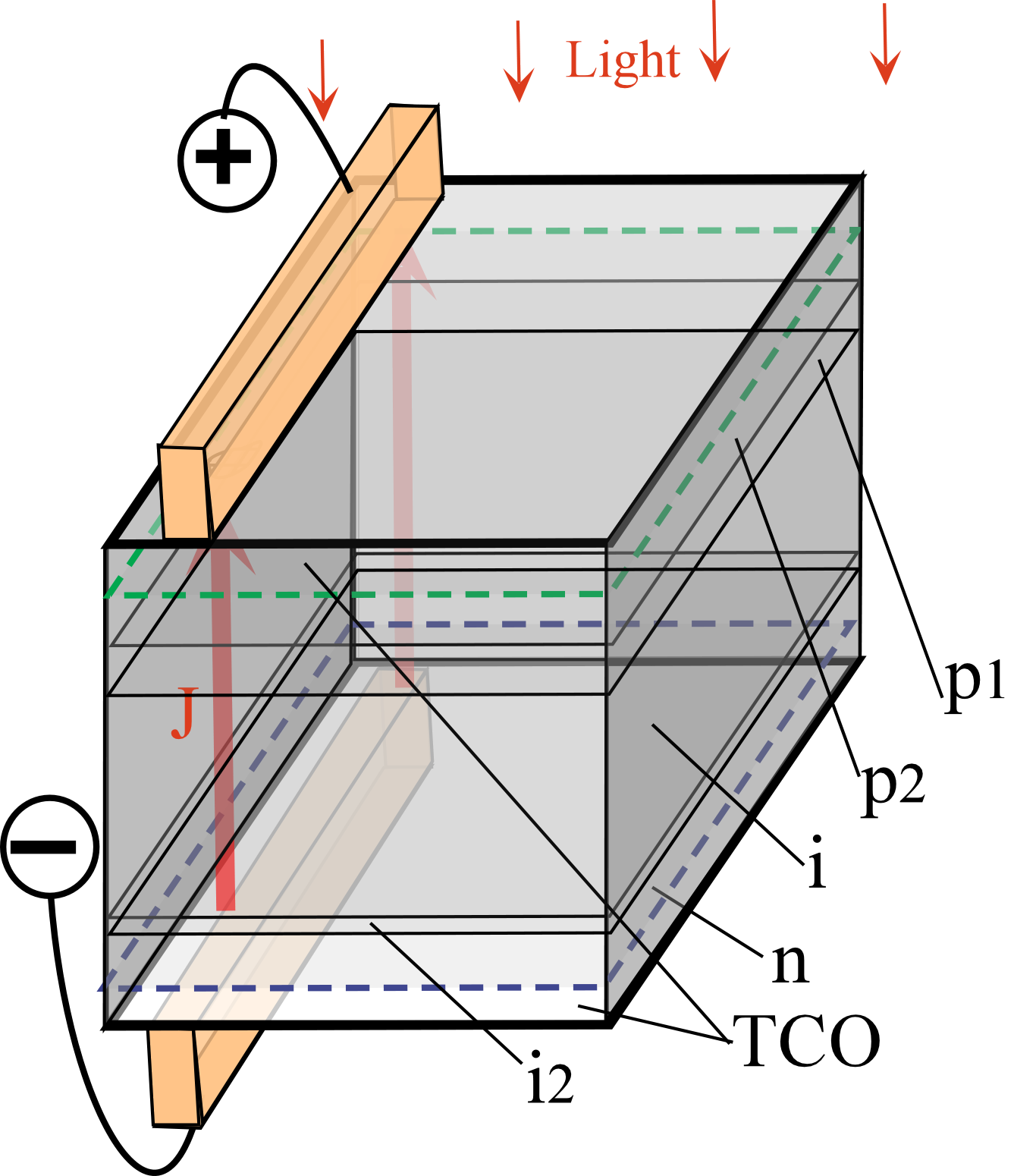} 
	\includegraphics[width=8cm]{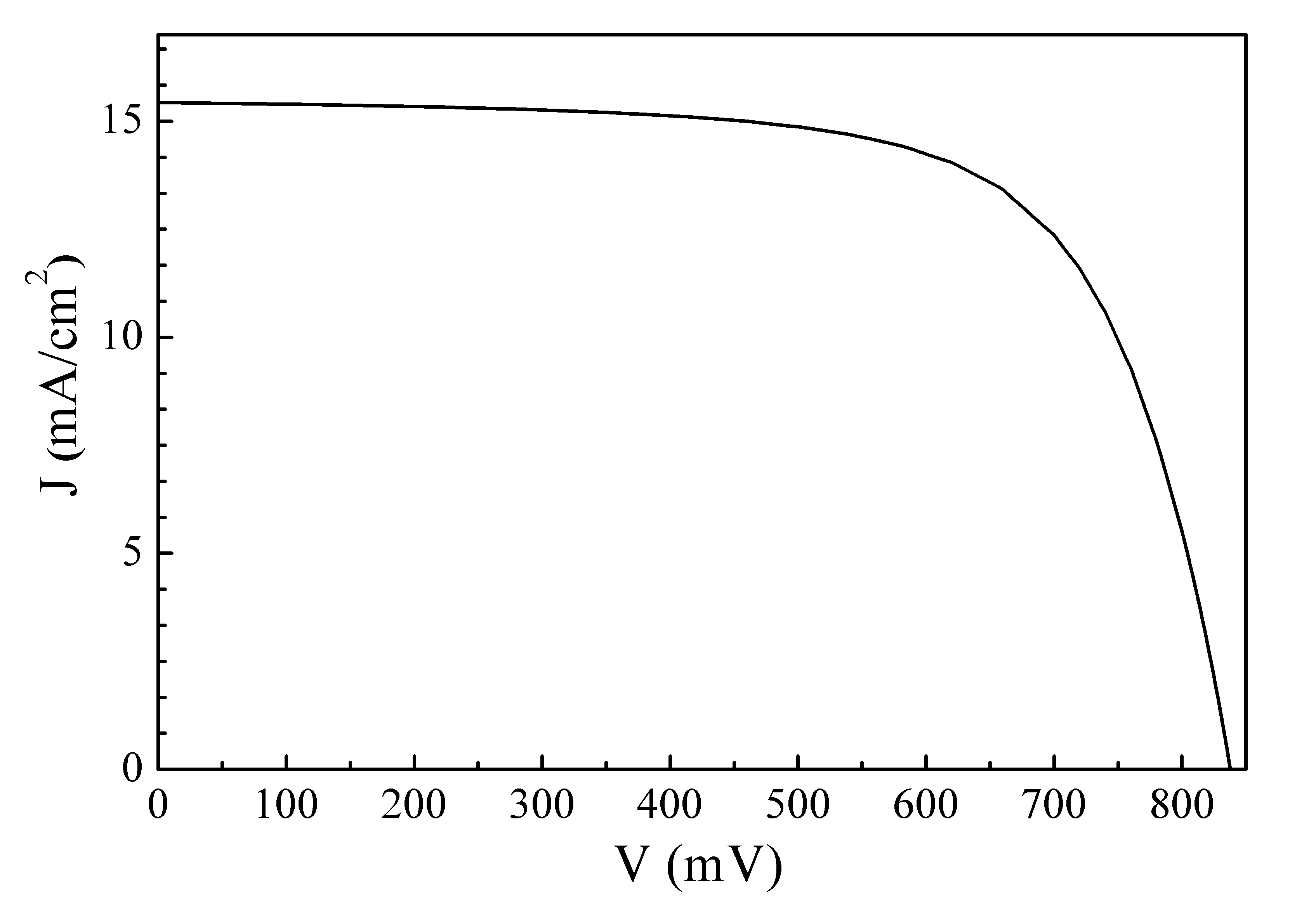}    
	\caption[(a)]{(a) Schematic structure of the device under investigation. (b) J-V characteristic curves of the  baseline simulated Cell-A. }
	\label{fig:fig1}
\end{figure}

\begin{table}
\caption{\label{jlab1}Parameters of the solar cell  layers (Cell-A) used in the simulation. MS-Schottky contact was used at the front TCO/ p-nc-SiO:H, TCO/ p-a-SiO:H, with surface recombination velocities of the electrons and holes as $10^{7}$ cm/s. For the other interfaces ‘drift-diffusion’ interface was 	used with zero interface charge density.}

\footnotesize
\begin{tabular}{@{}llllll}
	\br
	Parameters&p-ncSiO:H(p1)&p-a-cSiO:H(p2)&i-a-SiGe:H&i-a-Si:H&n-ncSi:H\\
	\mr
	Layer thickness,& 10&15&300&20&20 \\
	 d (nm)&   &  &  &  &  \\
	 \mr
	Dielectric constant & 11.9  & 11.9 & 11.9 & 11.9 & 11.9 \\
	\mr
	Electron affinity (eV) & 3.9  & 3.9 & 3.9 & 3.9 & 3.9 \\
	\mr
	Band gap (eV) &  2.08 & 2.08 & 1.58 & 1.68 & 1.68 \\
	\mr
	Conduction band  &   &  &  &  &  \\
	 density of states & $4\times 10^{20}$   & $4\times 10^{20}$ & $4\times 10^{20}$ & $4\times 10^{20}$ & $1\times 10^{20}$ \\
	 (cm$^{-3}$ )&   &  &  &  &  \\
	 \mr
	 Valence band  &   &  &  &  &  \\
	 density of states & $4\times 10^{20}$   & $4\times 10^{20}$ & $4\times 10^{20}$ & $4\times 10^{20}$ & $1\times 10^{20}$ \\
	 (cm$^{-3}$ )&   &  &  &  &  \\
	 \mr
Electron mobility& 0.2  & 0.2 & 2 & 1 & 10 \\
(cm$^2$ .volt$^{-1}$ .s$^{-1}$ )	&   &  &  &  &  \\
\mr
Hole mobility& 10  & 10 & 0.2 & 0.1 & 0.1 \\
(cm$^2$ .volt$^{-1}$ .s$^{-1}$ )	&   &  &  &  &  \\
\mr
	Doping&   &  &  &Dop.grad.  &  \\
	concentration & $1.6\times 10^{19}$   & $1.6\times 10^{18}$ &   & $(0 - 3) \times 10^{17}$ & $2.68\times 10^{18}$  \\
	(cm$^{-3}$ )&   &  &  &  &  \\
	\mr
	Layer density& 2.328  & 2.328 & 2.328 & 2.328 &2.328  \\
	(g.cm$^{-3}$ )&   &  &  &  &  \\
	\mr
	 Donor defect& $1\times 10^{16}$  & $1\times 10^{16}$ & $1\times 10^{16}$ &$1\times 10^{16}$  & $1\times 10^{16}$ \\
	 density (cm -3 )&   &  &  &  &  \\
	 \mr
	 Acceptor defect& 0  & 0 & $3\times 10^{16}$ &$1\times 10^{16}$  & $1\times 10^{16}$ \\
	density (cm -3 )&   &  &  &  &  \\
	\mr
	Defect type& Gaussian  & Gaussian & Gaussian & Gaussian & Gaussian \\

	\br
\end{tabular}\\
\end{table}

 \section{Results and Discussions}
 
The standard AM1.5G insolation (100 mW/cm$^{2}$	 of solar spectra) was used to estimate optical absorption of individual layers and J-V characteristic curves of the solar cells.  In order to get an idea of how the solar spectra is absorbed at different layers Fig. 2 shows the effective part of the AM1.5G spectra absorbed by different layers of the Cell-A. Here it can be seen that the top p-nc-SiO:H layer has high optical absorption in the shorter wavelength region. Although thickness of this layer is merely 10 nm, yet, because it is at the front side of the device, it absorbs significant amount of high energy photons. Generally, such an absorption is undesirable and several investigations were made to reduce such an absorption loss [10, 17, 18]. 

 \subsection{Optical Absorption}
 \subsubsection{Reference Cell-A}
 
The purpose of this p-nc-SiO:H layer was to improve device performance. But if two p-type layers are used, it inevitably raises total parasitic optical absorption at the p-type window layers. Although the optical absorption at the p2 layer is not very  high , yet it can be reduced to improve device performance. The simulated J-V characteristics curve obtained in the above figure (Fig. 1(b)) was obtained with both the p-type layers. It is expected that the device characteristics will improve by removing this p-nc-SiO:H layer. So this layer was removed for subsequent simulations. The resultant optical absorption spectra of the individual layers are shown in Fig. 3(a). 

\begin{figure}
	\centering
	\includegraphics[width=15cm]{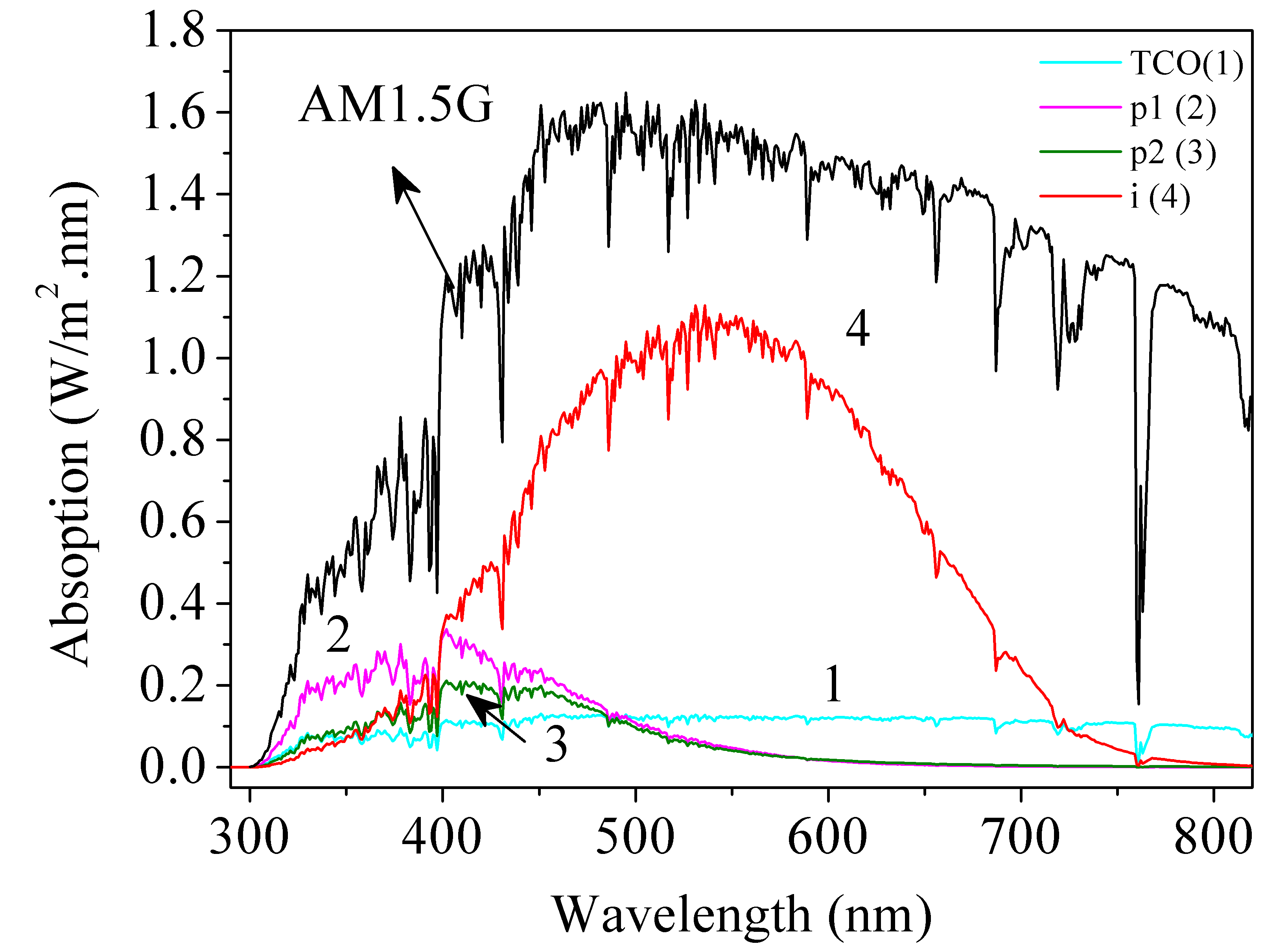} 
	\caption[(a)]{ Optical absorption spectra of different layers of the device, Cell-A. The top profile shows the AM1.5G spectrum. The numbers (1,2,3,4) in the legend correspond to the labeling of the traces. and ‘p1’, ‘p2’ and ‘i’ indicates p-nc-SiO:H, p-a-SiO:H, i-a-SiGe:H layers respectively.  Here the reported spectra was measured with 10 nm resolution. }
	\label{fig:fig2}
\end{figure}


Furthermore, AFORS-HET simulation shows that with a reduction in layer thickness of the layers p1, p2, the band discontinuity greatly increases, which is expected to have a deteriorating impact on device performance. On the other hand, it was also noticed that the discontinuity reduces with a thicker layers. It implies that with the p-type oxide double layer, there will be a lower limit of layer thickness that can be used to achieve better device performance. In a real device fabrication this may lead to an interpretation that a thinner window layer is not useful to improve PCE.

 \subsubsection{Modified Cell-A, without p-nc-SiO:H (Cell-B type)}
 
Figure 3(a) shows optical absorption of various layers of the cell (denoted as Cell-B) after the p1 (p-nc-SiO:H) layer was removed. The Fig 3(b) shows an increase in absorption by the i-a-SiGe:H active layer is taking place after removing the p-nc-SiO:H layer, as expected. Therefore, the electronic characteristics of the device are also expected to improve as well. 

\begin{figure}
	\centering
	\includegraphics[width=7cm]{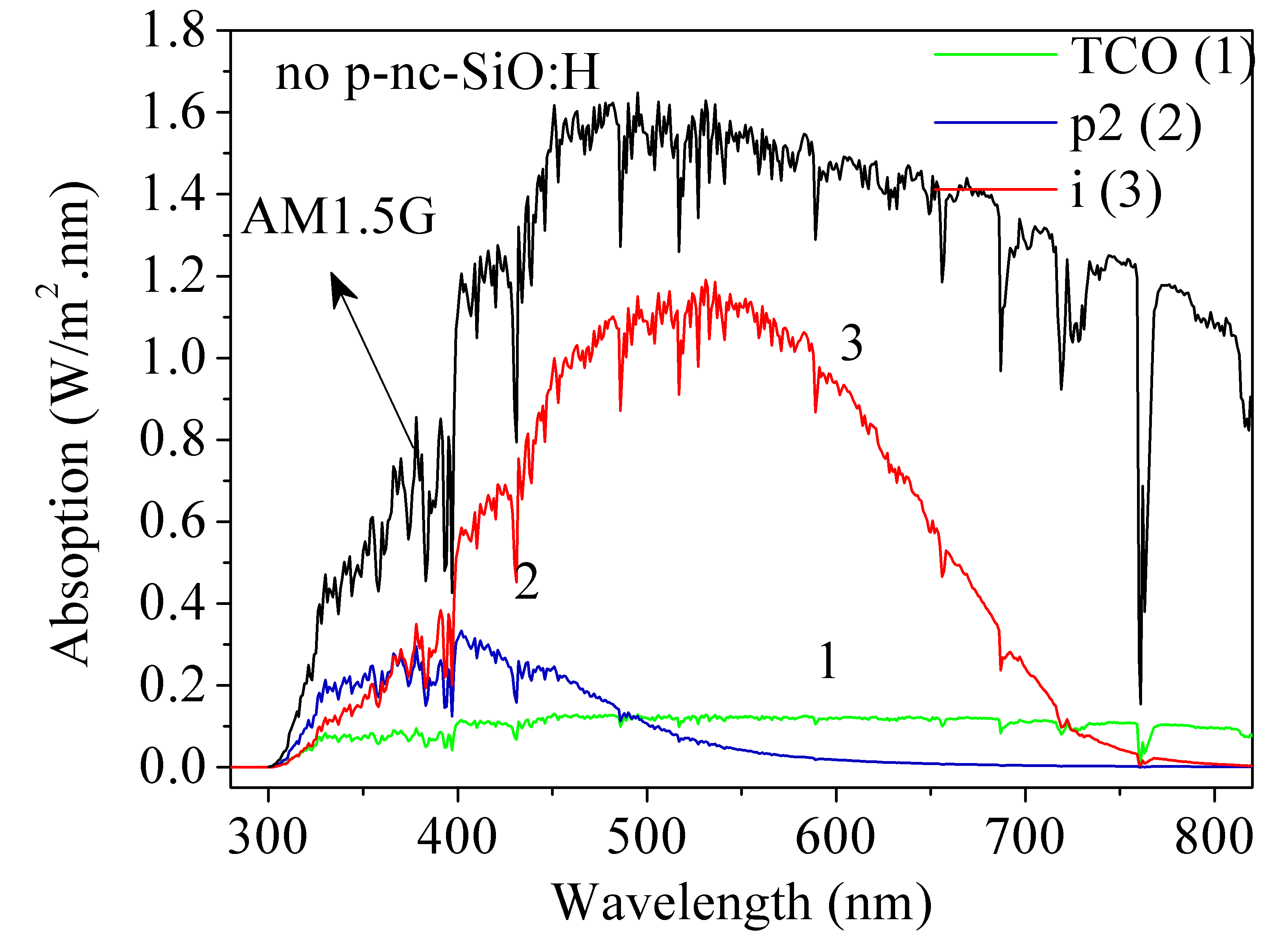} 
	\includegraphics[width=7cm]{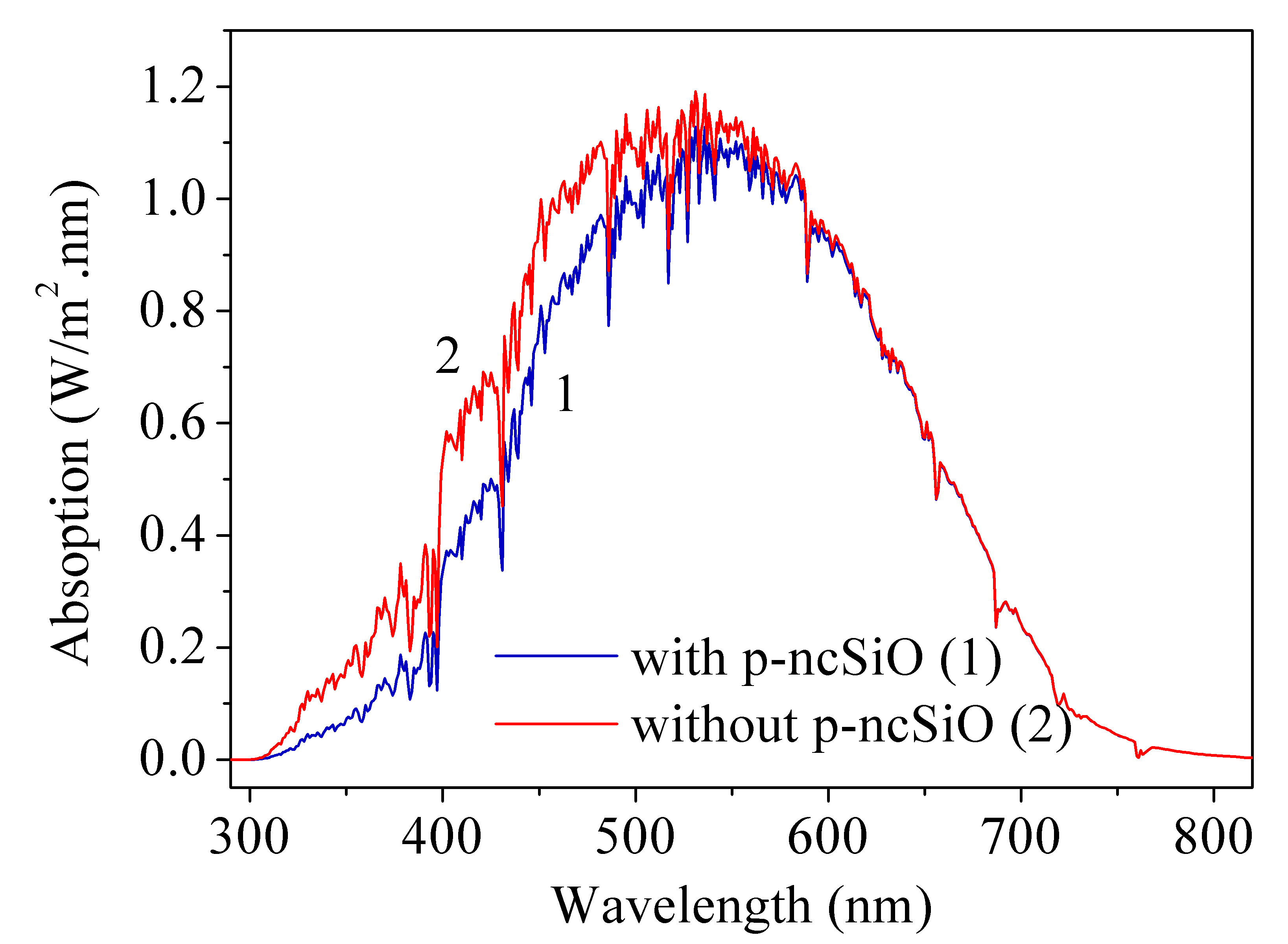}    
	\caption[(a)]{ (a) Optical absorption by the TCO, p2(p-a-SiO:H), intrinsic layer when the p-nc-SiO:H is absent, as in Cell-B. (b) Comparison of optical absorption by the active layer with (Cell-A) and without the p-nc-SiO:H layer (Cell-B). }
	\label{fig:fig3}
\end{figure}


 \subsubsection{$d_{i}$ variation (Cell-B type)}
 
One of the most popular and simple approach to improve device performance is to find optimum thickness of the active layer. Generally it is thought that by increasing thickness of the active layer can raise short circuit current density ($J_{sc}$) of the cell. This is generally true if defect density of the active layer is low. However, the $J_{sc}$ is not the only parameter used in estimating PCE, as the PCE is estimated by using product of current density and voltage at the maximum power point (PmaxJ, PmaxV respectively). So raising $J_{sc}$ not necessarily can improve device performance if fill factor (FF) and output voltage across the  device degrades. To start with the related simulation, the variation in light absorption at the active layer for its various thicknesses ($d_{i}$) were obtained and is shown in Fig. 4(a):

\begin{figure}
	\centering
	\includegraphics[width=7cm]{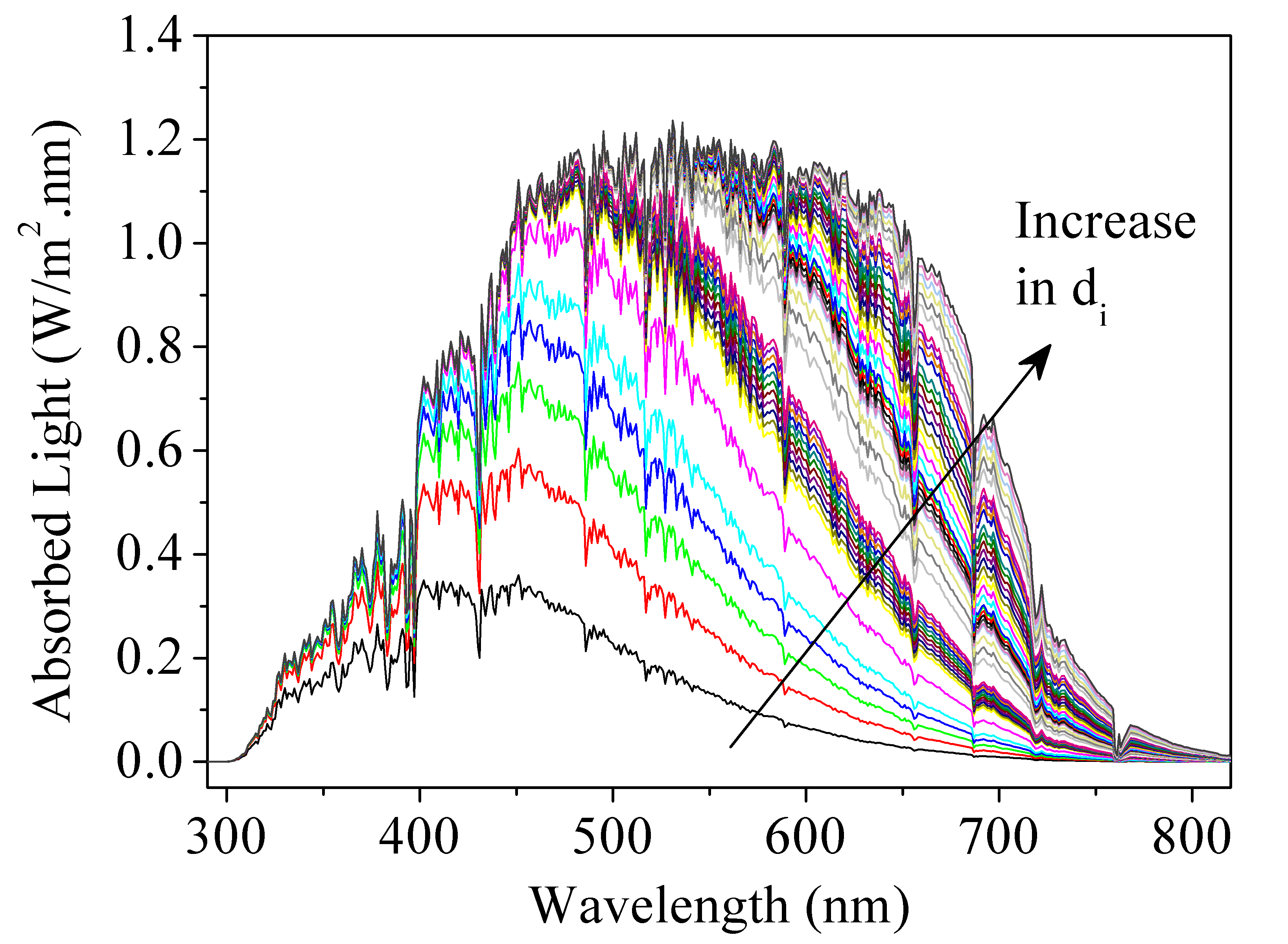} 
	\includegraphics[width=7cm]{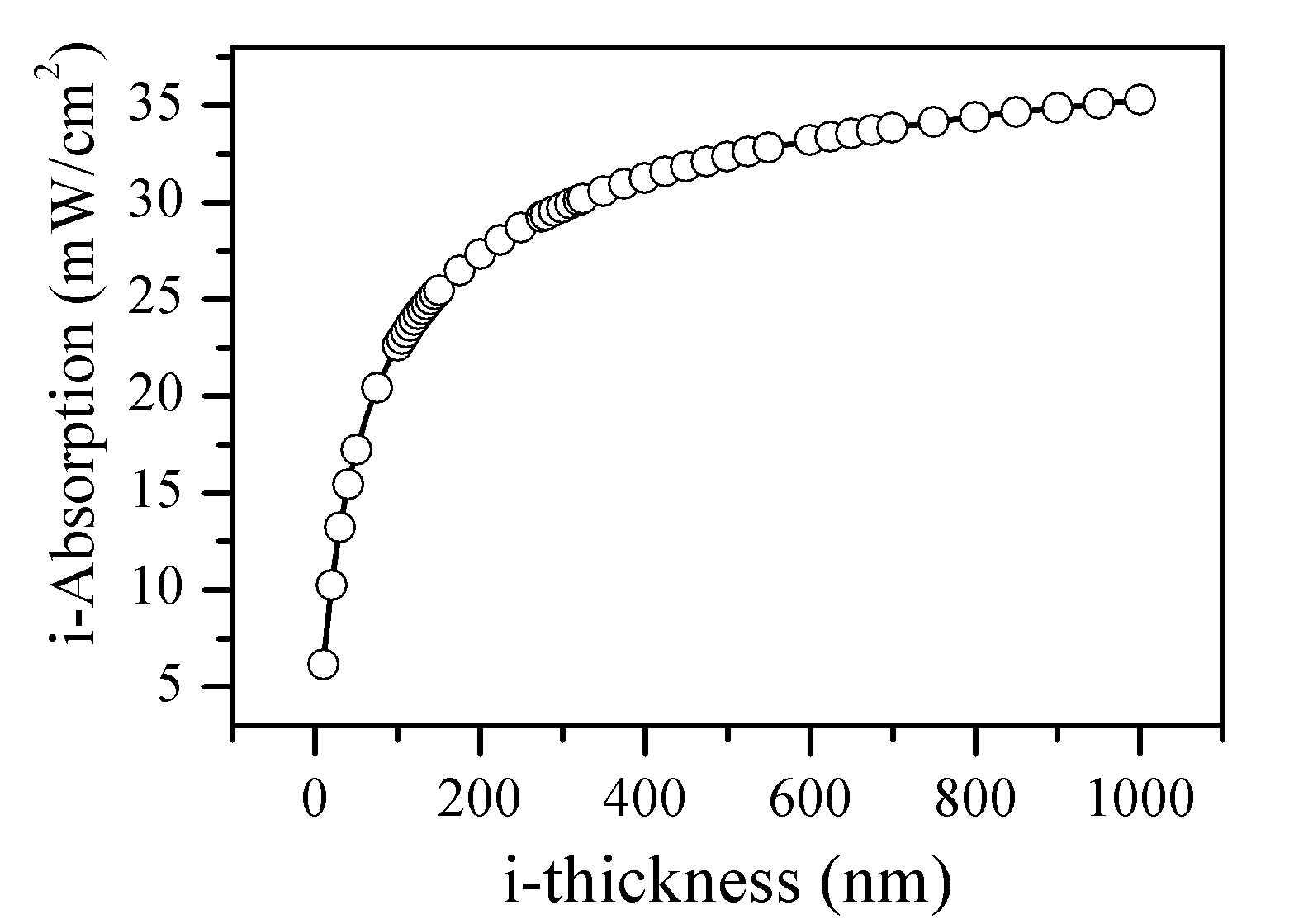}    
	\caption[(a)]{ (a) Optical absorption by the i-a-SiGe:H layer for its various thicknesses ($d_{i}$) of active layer of the Cell-B, as (10, 20, 30, 40, 50, 75, 100, 105, 110, 115, 120, 125, 130, 135, 140, 145, 150, 175, 200, 225, 250, 275, 280, 290, 300, 310, 320, 325, 350, 375, 400, 425, 450, 475, 500, 525, 550, 600, 625, 650, 675, 700, 750, 800, 850, 900, 950, 1000) nm. (Here, and the subsequent, estimates are made after removing the p-nc-SiO:H layer from the device structure.). Thickness of p-type layer was 7 nm. (b) Integrated optical absorption in the active layer denoted as i-Absoprion for various thickness of theactive layer, denoted as i-thickness. }
	\label{fig:fig4}
\end{figure}


This Fig. 4(a) shows that by increasing thickness of the active layer the wavelength response of the device improves along with the increase in the magnitude of peak absorption, Shift of the peak towards longer wavelength and an overall increase in light absorption.  It indicates that the long
wavelength response of the device changes significantly with the change in the $d_{i}$ . Integrated light absorption, as shown in Fig. 4(b) determines $J_{sc}$. Here it is interesting note the region where light absorption increases sharply with the increased thickness, which is below 100 nm. Furthermore, it is worth mentioning that here the optical absorption does not include the contribution from the i/n buffer layer, which is also an intrinsic layer and will contribute to the generated current of the device. For evaluating electronic properties of the device, the contribution from the buffer layer is taken into account.

 \subsubsection{$d_{p}$ variation (Cell-B type)}
 
We further investigated the change in parasitic optical absorption due to change in the thickness of the p-type window layer ($d_{p}$) of Cell-B. Fig. 5(a) shows the optical absorption by the p-type layer when its thickness was varied from 1 nm to 25 nm. The p-type layer absorbs short wavelength part of the spectra or high energy photons. As a result, the available light to the intrinsic layer will also change and hence there will be a change in light absorption by the active layer. The spectra of absorbed light by the active layer are shown in Fig. 5(b). It shows that by changing width of the p-type window layer the blue response (short wavelength response) of the device changes significantly, whereas the long wavelength response remain mostly unchanged. Integrated light absorption at the p-type and i-type layers, as obtained from the Fig. 5(a) and 5(b), are shown in Fig. 5(c). It further highlights that the reduced parasitic absorption at the p-layer leads to an increased absorption at the i-layer, while sum of these two absorptions remain nearly constant to $\approx$32 mW/cm2. Therefore this is intuitively clear that higher optical absorption at the active layer can be achieved with a thinner p-layer.

\begin{figure}
	\centering
	\includegraphics[width=7cm]{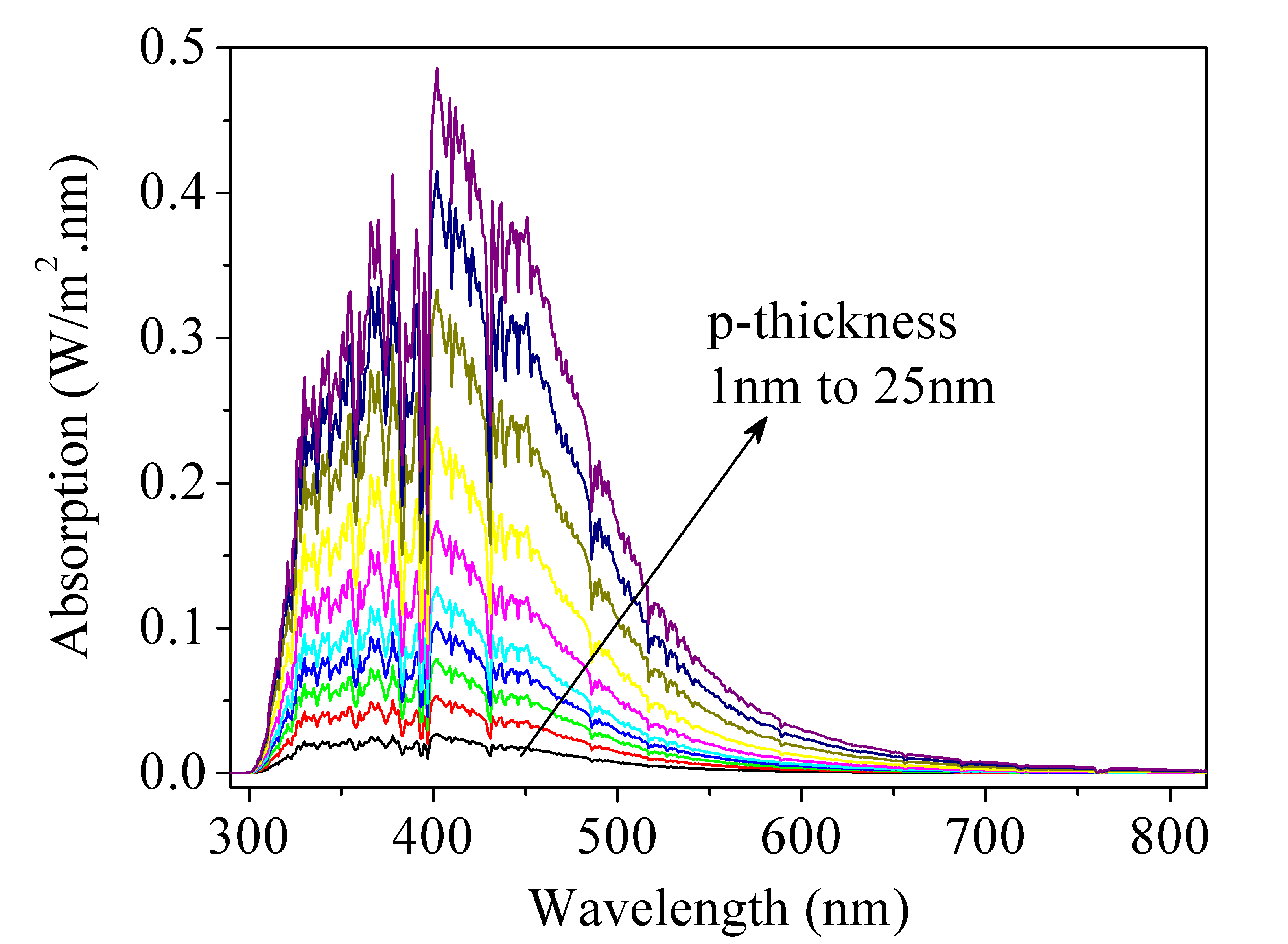} 
	\includegraphics[width=7cm]{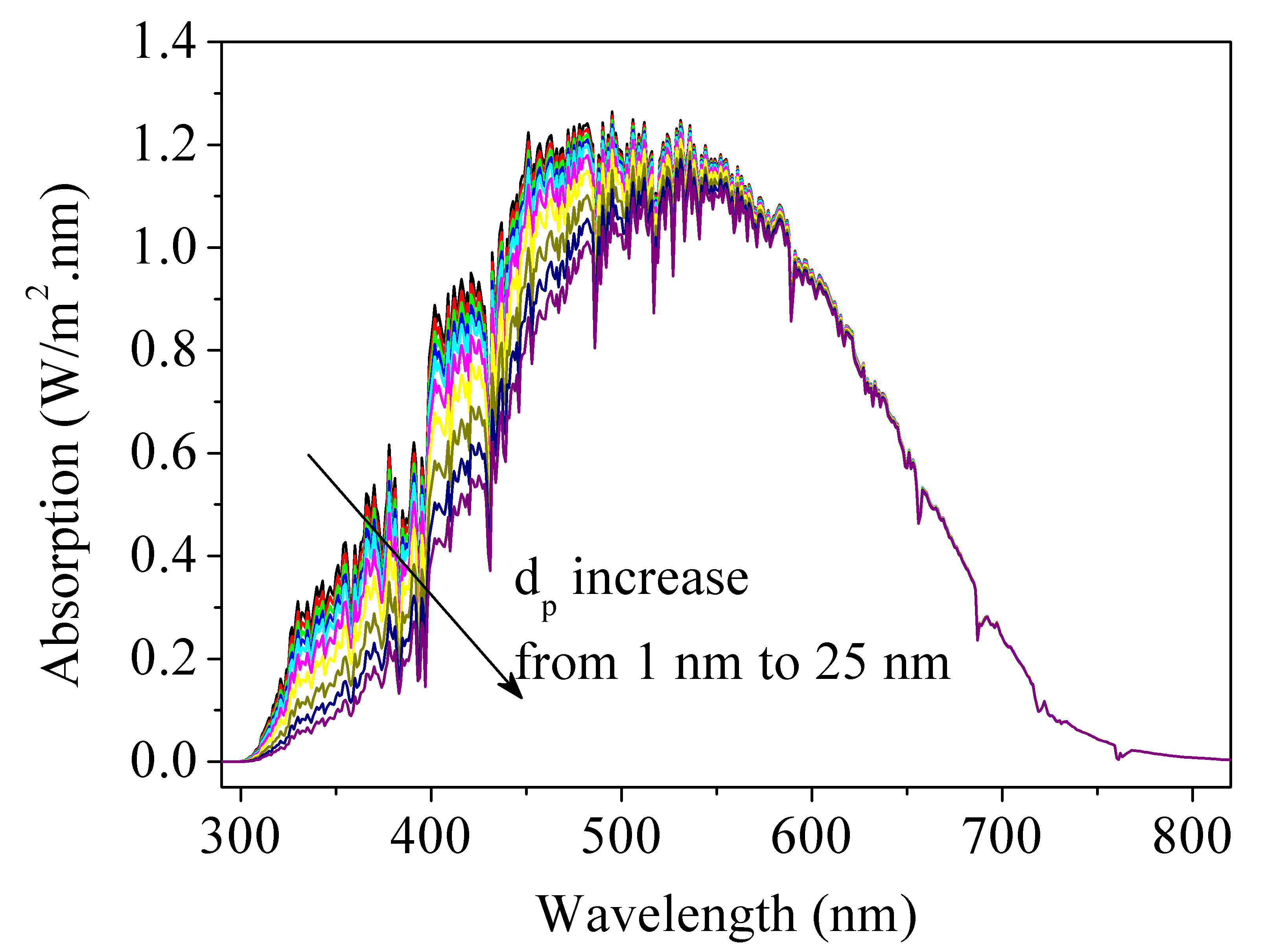}    
	\includegraphics[width=7cm]{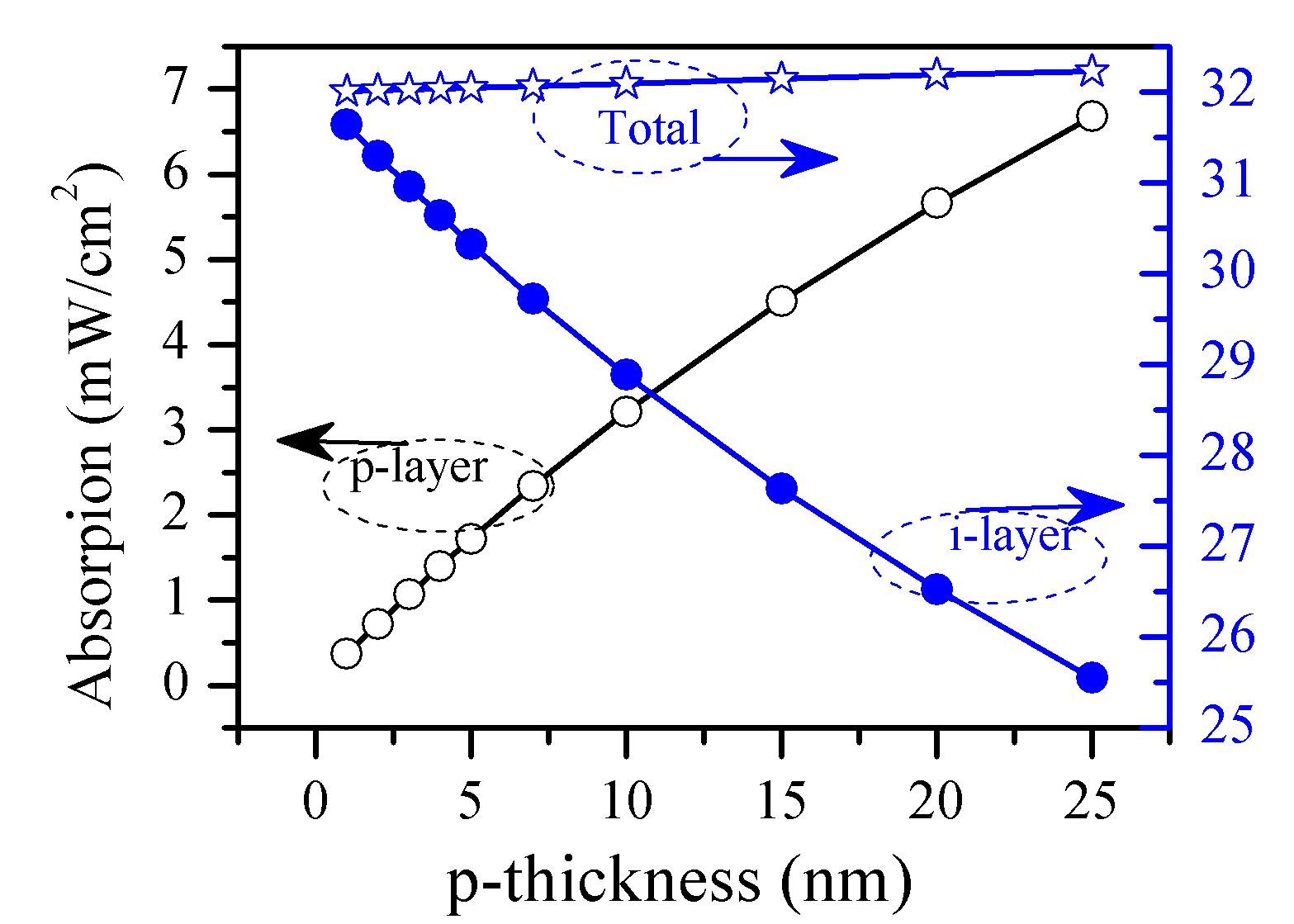}    
	\caption[(a)]{ (a) Optical absorption by the p-a-SiO:H layer, for its thicknesses ($d_{p}$) as 1, 2, 3, 4, 5, 7, 10, 15, 20, 25 nm and a corresponding optical absorption by the active layer as shown in (b). Here the active layer thickness ($d_{i}$) is 100nm. (c) Estimated total energy absorbed by the p-type and i-type layer, and their sum, denoted as ‘p-layer’, ‘i-layer’ and ‘Total’ respectively in the figure.  }
	\label{fig:fig5}
\end{figure}


Reducing optical absorption at the p-layer can raise absorption in the i-layer ad hence raise PCE of the device. A simple way to reduce the absorption at the p-layer is to reduce its thickness.

 \subsection{Electrical Characteristics}
 
 \subsubsection{$d_{i}$ variation, without p-nc-SiO:H (Cell-B type)}
 
As the photo generated current in a solar cell is directly related to absorbed light by the active layer, it is expected to change the J-V characteristic curve of the device. Fig. 6 show J-V characteristic curves of the devices when the p-nc-SiO:H layer is present, Fig. 6(a), and when it is removed Fig. 6(b) and the thickness of the active layer varied in a wide range, from  20 nm to 1$\mu$m. It becomes visually clear that the short circuit current density ($J_{sc}$) mostly increases with thickness of active layer, but fill factor (FF) reduces significantly. Quantitative estimates of the solar cell parameters are shown in Fig. 7. 

\begin{figure}
	\centering
	\includegraphics[width=7cm]{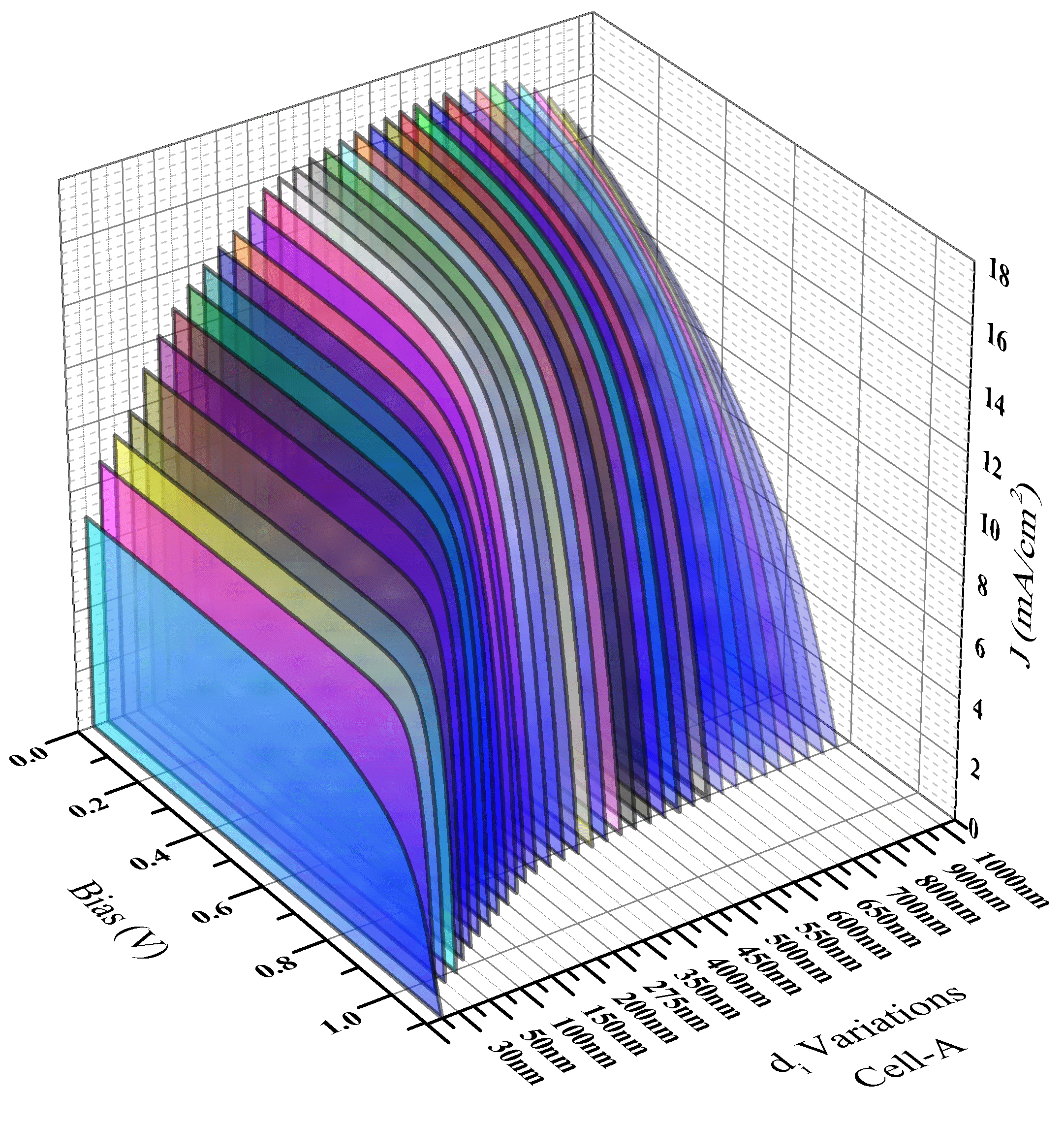} 
	\includegraphics[width=7cm]{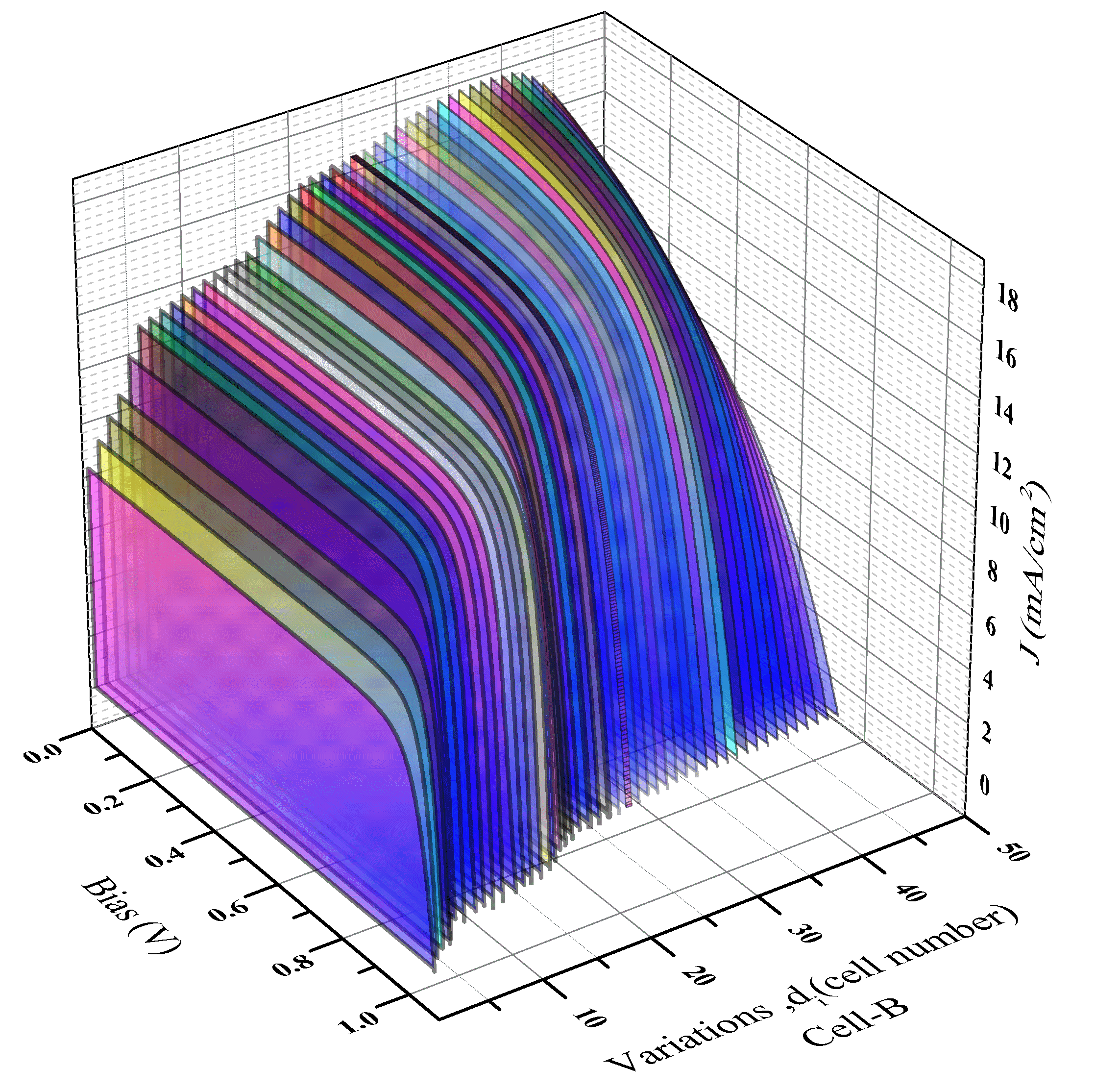}    
	\caption[(a)]{ J-V characteristic curves of the cells (a) Cell-A type, (b) Cell-B type, with various thicknesses of the active layer, thicknesses as mentioned in Fig. 4, $d_{p1}$ = 15nm. }
	\label{fig:fig6}
\end{figure}


The characteristic features of the J-V curve for the Cell-A and Cell-B,  is shown in Fig 6 and Fig. 7. The open circuit voltage ($V_{oc}$) decreases sharply when active layer thickness ($d_{i}$) was varied from 20 nm to 300 nm. With a further increase in the thickness beyond 300 nm, the $V_{oc}$ decreases at a slower rate. Similarly, the increase in the $J_{sc}$ is sharper until 300 nm, but the rate of increase slows down  within 300 nm to 800 nm and it starts reducing after $d_{i} > 800$ nm. The FF continuously decreases with the increase in the $d_{i}$. It implies that the $V_{oc}$ and FF decreases monotonously with the increase in the $d_{i}$ which can be interpreted as a result of total number of defects in the active layer. On
the other hand, while the $J_{sc}$ increases primarily because of increased optical absorption, but it also decreases for $d_{i} > 800$ nm. As a result the maximum PCE was obtained with 125 nm thick active layer for Cell-A and 130 nm for Cell-B, which are much lower than that the active layer thickness ($\approx 300$ nm) conventionally used, yet the PCE of this device is higher than that of the starting  Cell-A  ($d_{i} = 300$ nm).

\begin{figure}
	\centering
	\includegraphics[width=15cm]{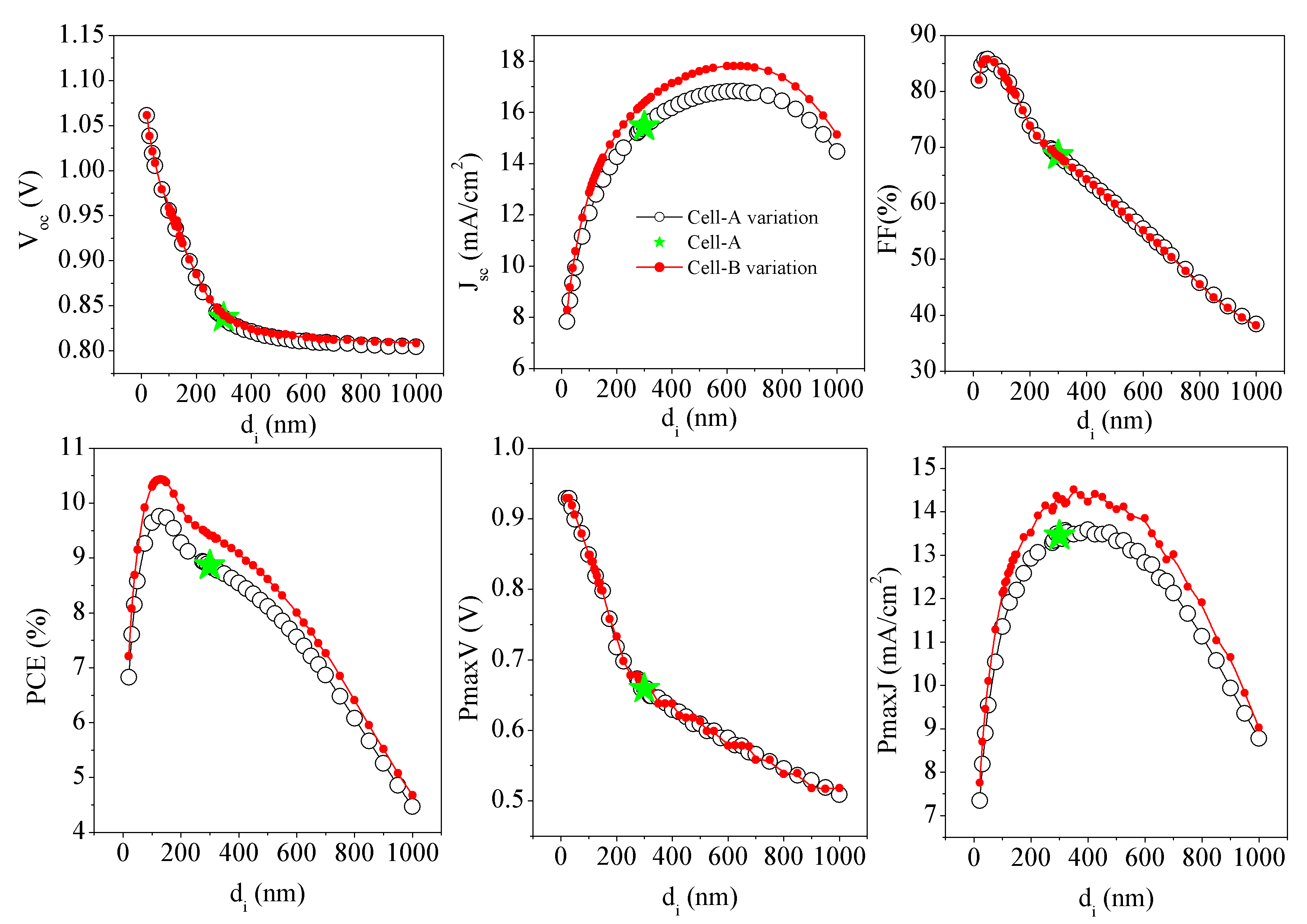} 
	\caption[(a)]{ Solar cell characteristic parameters  (for Cell-A,B), extracted from the J-V curves in Fig. 6. The respective value of the reference cell is shown as star-symbol. The maximum PCE of the Cell-B type was observed at $d_{i} \approx 125$ nm. }
	\label{fig:fig7}
\end{figure}


The 20 nm thick active layer may appear impractical, but the result demonstrates that a thinner device can exhibit a very high $V_{oc}$ and FF. In this context, it is to be noted that device a thinner layer ($\approx 3$ nm thick layer) was also investigated earlier [19]. The reason can be the average number of defect centers encountered by photo generated charge carriers. Defect density ($N_{dd}$ ), which is the number of defects per
unit volume, determines SRH recombination rate ($R_{SRH}$ ) so the total recombination of the photo generated carriers ($R_{ec}$ ) will be determined by the product of $R_{SRH}$ and volume ($V_{olume}$ ) of active layer (where$V_{olume}= d i *A$, A is surface area). Therefore

\begin{equation}
	R_{ec} = R_{SRH}V_{olume}N_{dd} = k_{1} d_{i}
\end{equation}

Here $k_{1} = R_{SRH}V_{olume}N_{dd}$, $N_{dd}$ is a constant when the recombination rate, $A$ is surface area and defect density are constant. This indicates that the total recombination of photo generated carriers is directly proportional to thickness of active layer when the $d_{i}$ is varied.

Interestingly, the estimated PCE is highest ($9.76\%$) with  $d_{i}$ ($\approx$125nm)for Cell-A and $10.43\%$ with 130 nm active
layer for Cell-B. Practically the device can be made further thinner. With the  $d_{i} =  100$ nm the PCE = $9.64\%$ which is 0.12 percentage point lower for a $20\%$ reduction in $d_{i} $ from 125 nm to 100 nm or to 1/3-rd of its original value (from 300 nm to 100 nm). From the cost of material point of view, the $d_{i} $ = 100 nm may be a preferable choice, as processing time may be an important parameter [20] for commercial production. It is also to necessary to mention that the PCE of the device (Cell-A) with 300 nm active layer increases marginally, from $8.87\%$ to $9.41\%$ by removing the p-nc-SiO:H layer, although Fig. 3(b) shows a significant increase in the optical absorption by removing this layer. It may be primarily because the defects in the active layer, so the additional photo generated carriers are mostly lost by recombination to travel across the 300nm thick active layer. Such an excess recombination loss can be reduced, by using a thinner active layer. As a result devices with 125 nm or 130 nm thick active layer shows better performance.

\subsection{ $d_{p}$ variation, without p-nc-SiO:H (Cell-B type)}

\begin{figure}
	\centering
	\includegraphics[width=8cm]{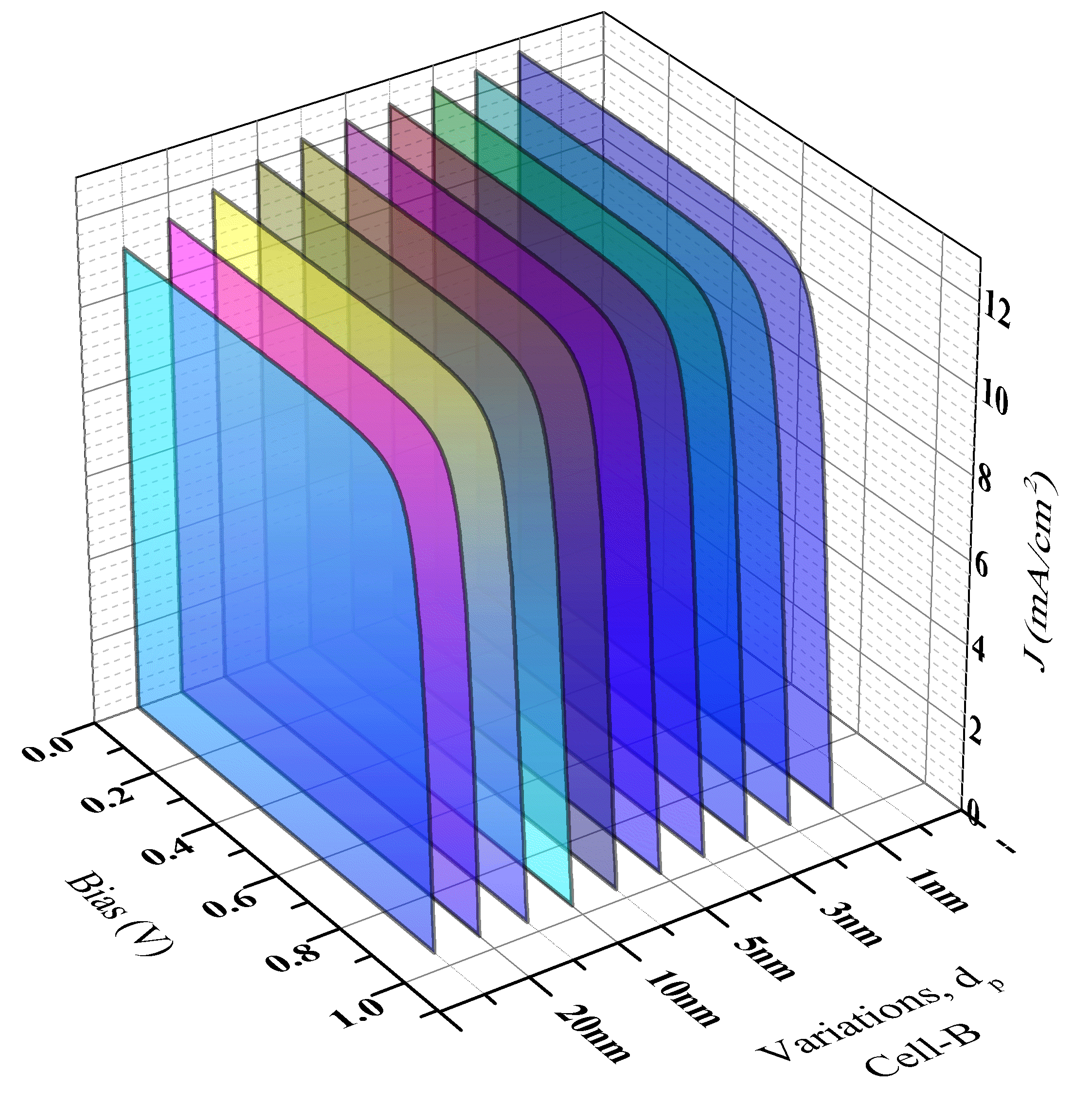} 
	\caption[(a)]{ J-V characteristic curves of the cells with various thicknesses of the p-a-SiO:H layer, thicknesses as mentioned in Fig. 5(b), with thin active layer, $d_{i} $ = 100 nm. }
	\label{fig:fig8}
\end{figure}


Based on the above discussion, the J-V characteristic curves, as shown in Fig. 8 and device parameters as shown in Fig. 9 were obtained for the $d_{i} $ = 100 nm. The $J_{sc}$ increases from $\approx$ 11 mA/cm$^{2}$ to $\approx$ 12.8 mA/cm$^{2}$ when the $d_{p}$ was reduced from 25 nm to 1nm. 

The $V_{oc}$ and FF of all the cells remain higher than the reference cell (star symbol  in Fig. 9). Although $J_{sc}$ remained lower than the reference cell yet, the PCE came out to be higher than when the $d_{p}$ <20nm. The result is interesting because it shows that obtaining a higher $J_{sc}$ may not always lead to a higher PCE. 

The effect of device structure on the short and long wavelength response may be comparable to that expected in a device with light trapping scheme. The use of a textured surface may have a possibility to introduce additional defects in the deposited film [21]. This will invite two different issues, one is that optimized device thickness inversely depends on defect density (as discussed later in this article), the other is that the surface texture may introduce additional non-uniformity of the film over the surface texture, whereby depositing ultrathin layer may not be possible to result a better device performance.

\begin{figure}
	\centering
	\includegraphics[width=12cm]{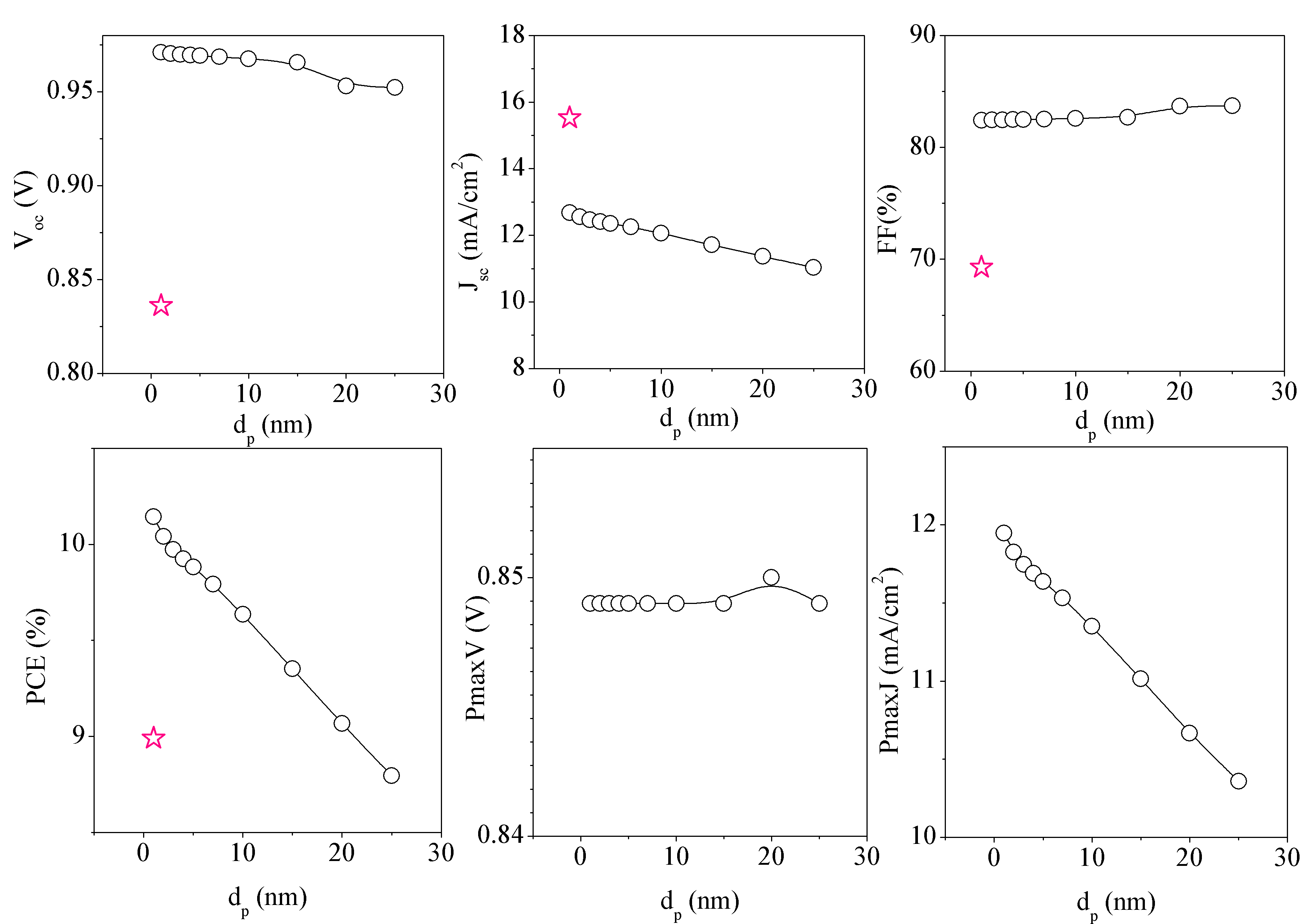} 
	\caption[(a)]{ Solar cell characteristic parameters extracted from the J-V curves in Fig. 8. The respective value of the reference cell is shown as star-symbol.  }
	\label{fig:fig9}
\end{figure}


 \subsection{$d_{i}$ Variation of Reference Cell type with $N_{d}= 6\times 10^{15}$ cm$^{-3}$}
 
The Cell-A and Cell-B had active layer defect density ($N_{d}$ ) of $6\times 10^{16}$ cm$^{-3}$ . Here, in the Cell-C type, the
defect density is reduced by one order. We denote the cells derived from Cell-A as Cell-AC, and those derived from Cell-B as Cell-BC1.  Defect density of the active layer generally has a profound effect on device performance. Therefore, lower defect density of active layer is always preferable.

\begin{figure}
	\centering
	\includegraphics[width=7cm]{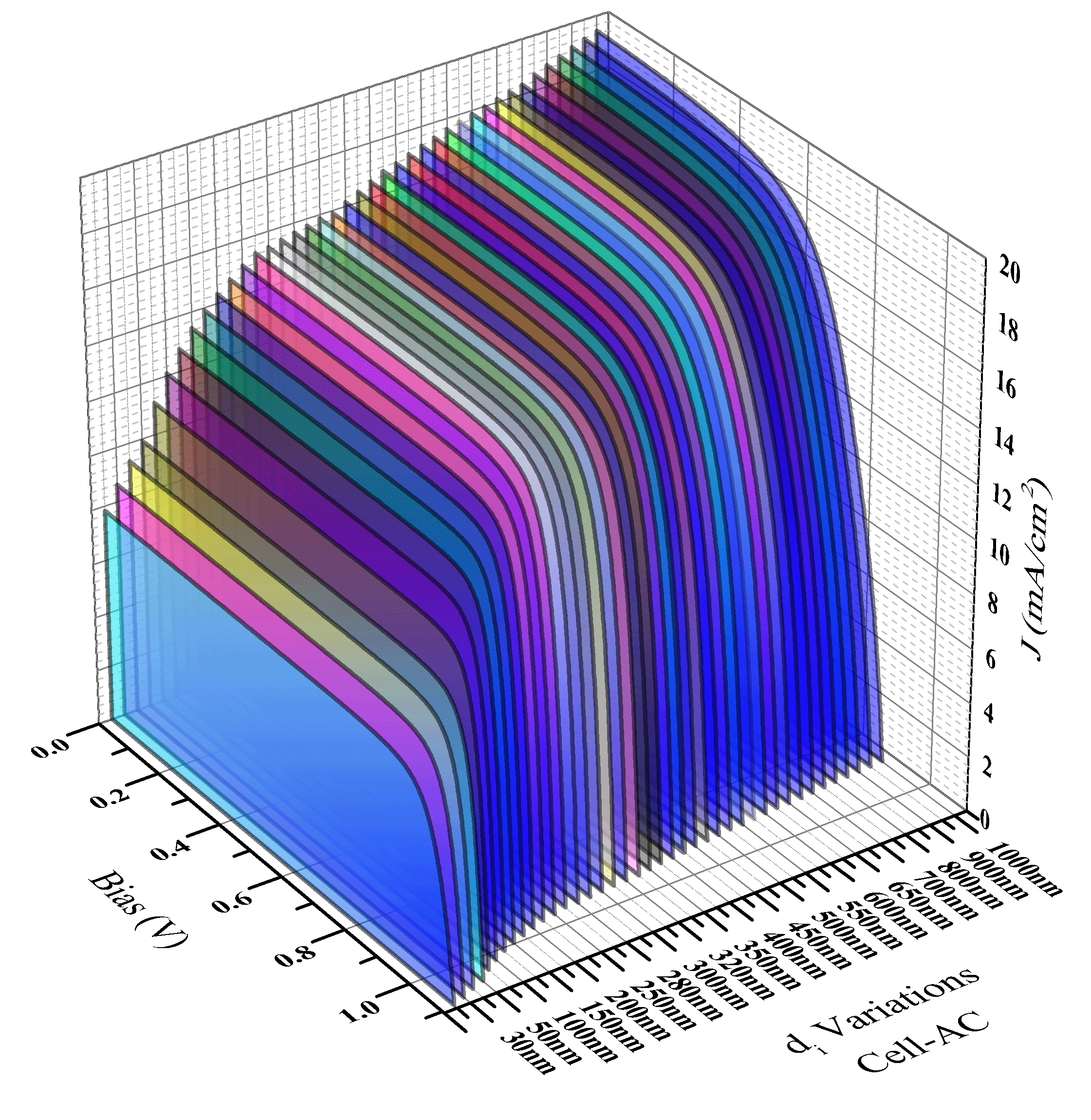} 
	\includegraphics[width=7cm]{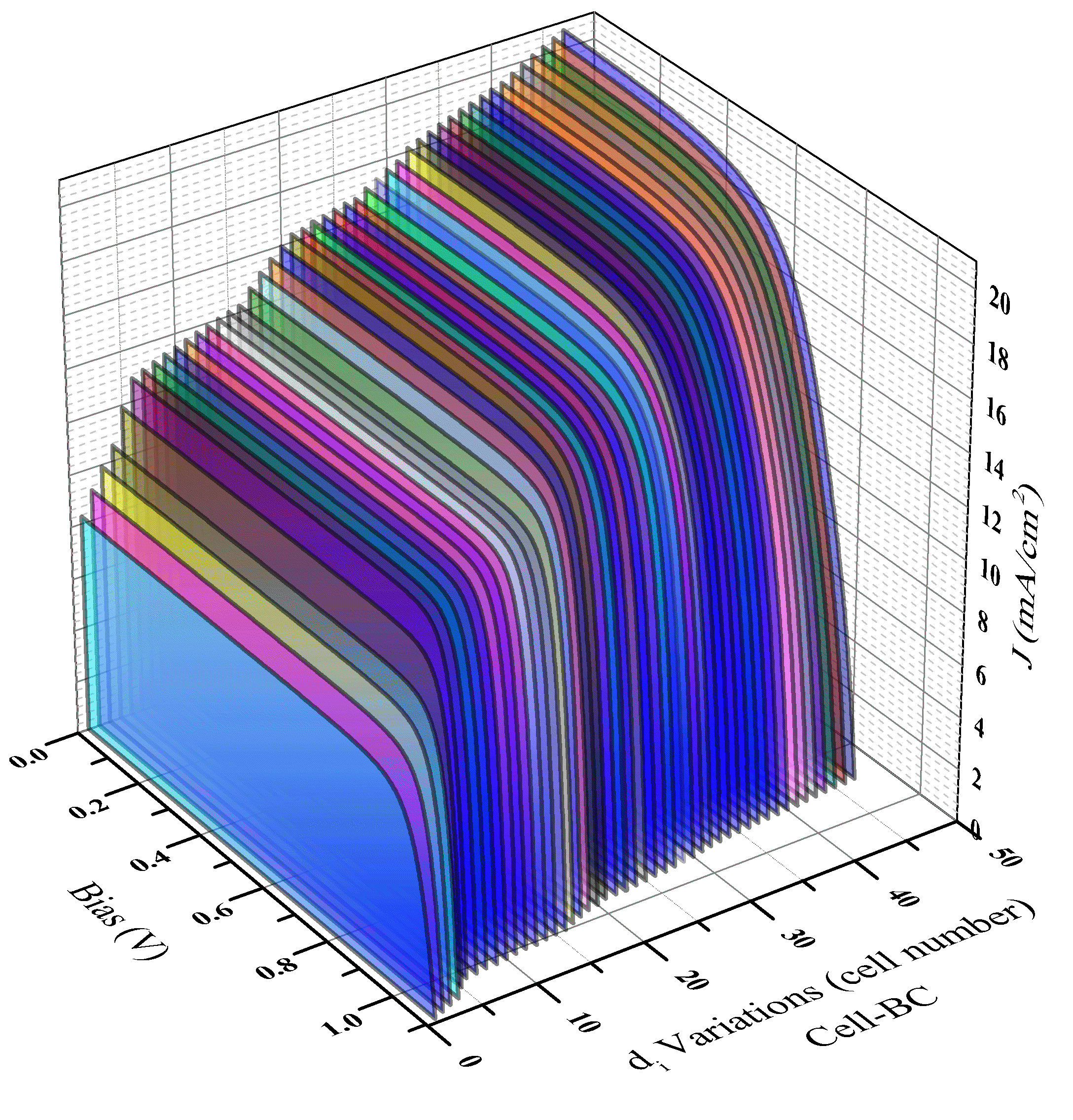}    
	\caption[(a)]{  J-V characteristic curves of the (a)cells (Cell-AC type), (b) Cell-BC1,similar to that of Fig. 6, except that here total defect density is reduced from $6\times 10^{16}$ cm$^{-3}$ to $6\times 10^{15}$ cm${-3}$  }
	\label{fig:fig10}
\end{figure}


\begin{figure}
	\centering
	\includegraphics[width=14cm]{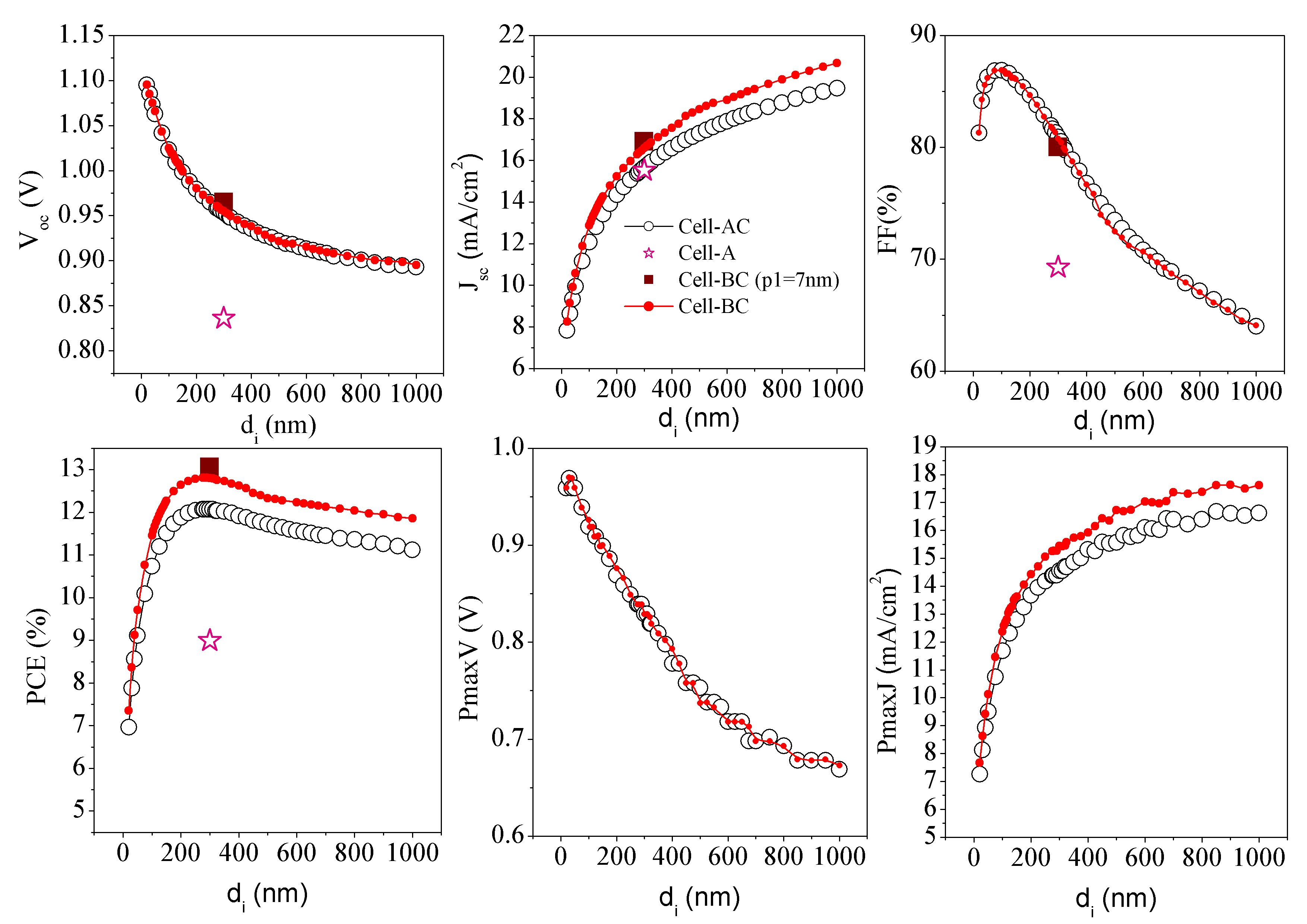} 
	\caption[(a)]{ Solar cell characteristic parameters extracted from the J-V curves from Fig. 10. Here the parameters are compared for devices  without the p-nc-SiO:H, but with 15 nm thick p-a-SiO:H (Cell-BC1), and with both the p-layers (Cell-AC). In comparison to the Fig. 6, 7, all other parameters are the same except the defect density ( defect density was reduced by one order of magnitude).  The respective value of the reference cell (Cell-A) is shown as star-symbol. The filled-square symbol corresponds to $d_{i} =300$nm, without p-nc-SiO:H, and 7 nm thick p-a-SiO:H, with defect density $6\times 10^{15}$ cm$^{-3}$ (as Cell-BC2). }
	\label{fig:fig11}
\end{figure}


\begin{figure}
	\centering
	\includegraphics[width=7cm]{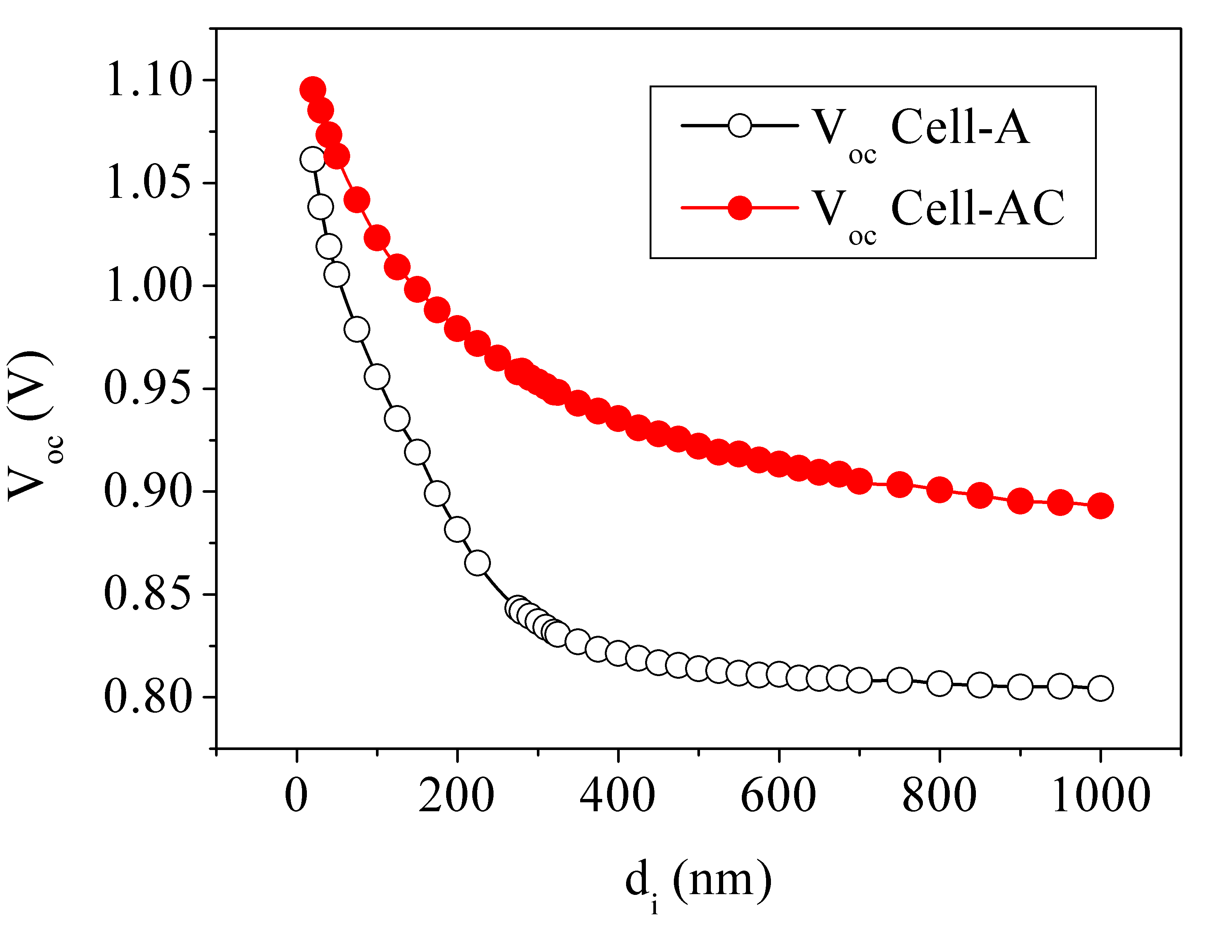} 
	\includegraphics[width=7cm]{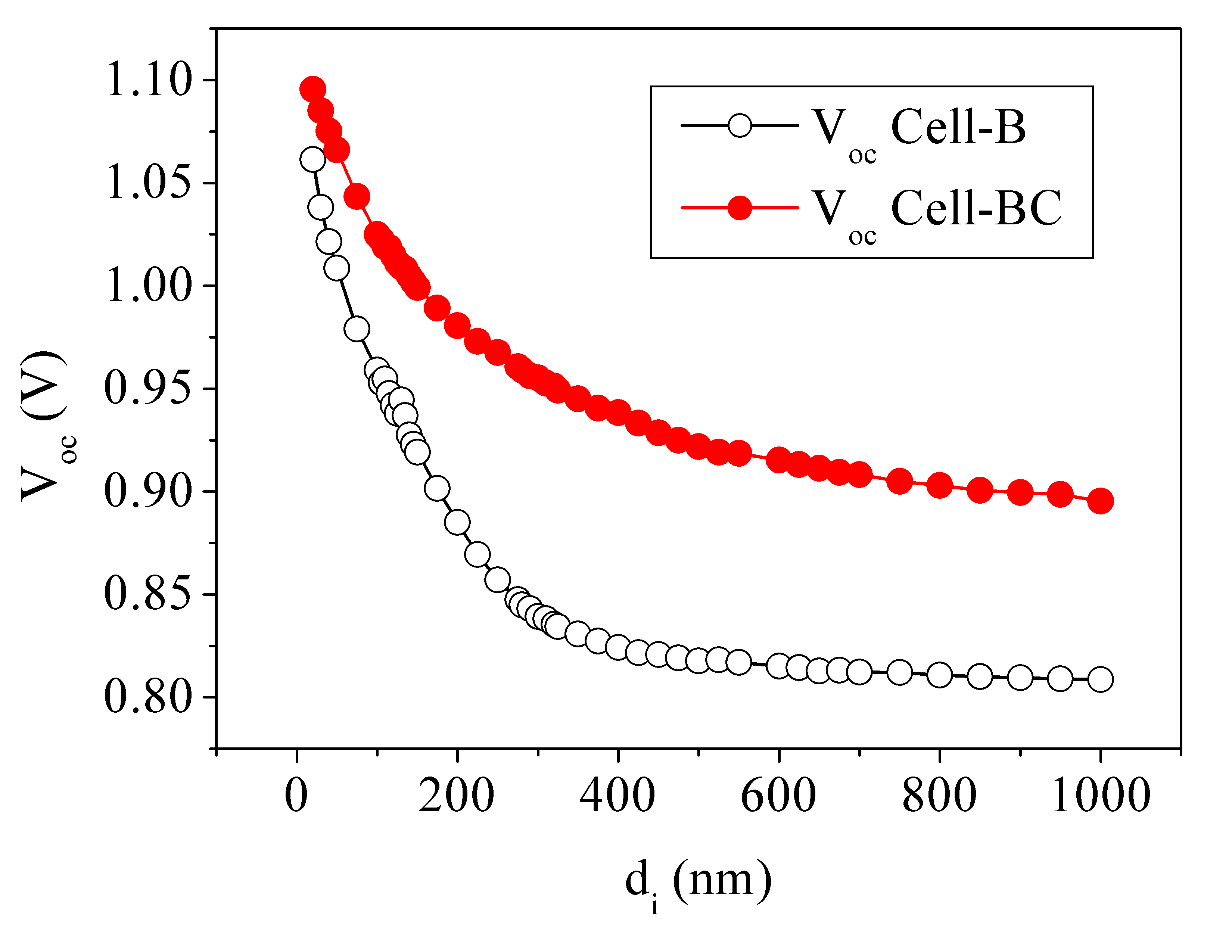}    
	\caption[(a)]{ Variation of the open circuit voltages of the (a) Cell-B and (b) Cell-BC, taken from Fig. 7 and 11 respectively. }
	\label{fig:fig12}
\end{figure}

\begin{figure}
	\centering
	\includegraphics[width=7cm]{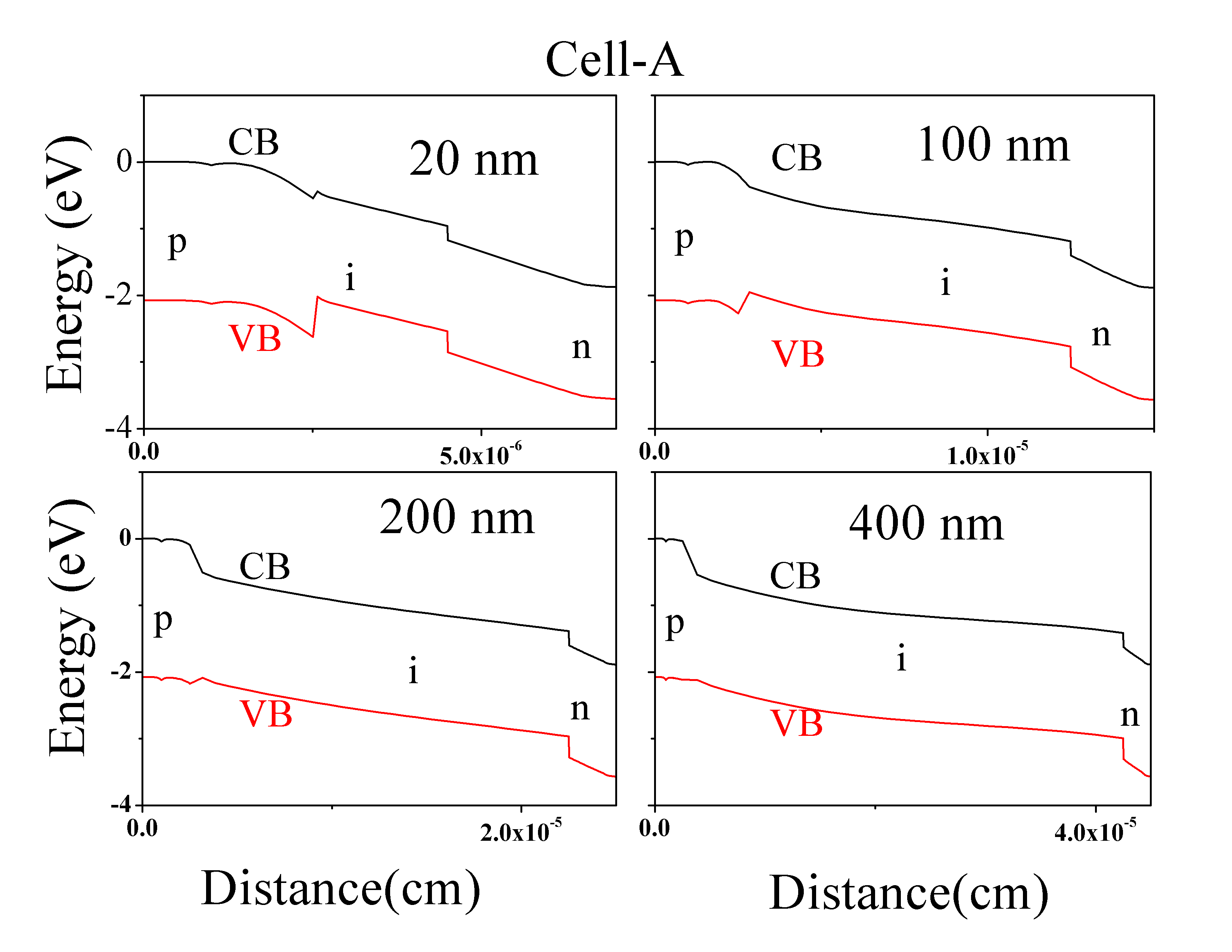} 
	\includegraphics[width=7cm]{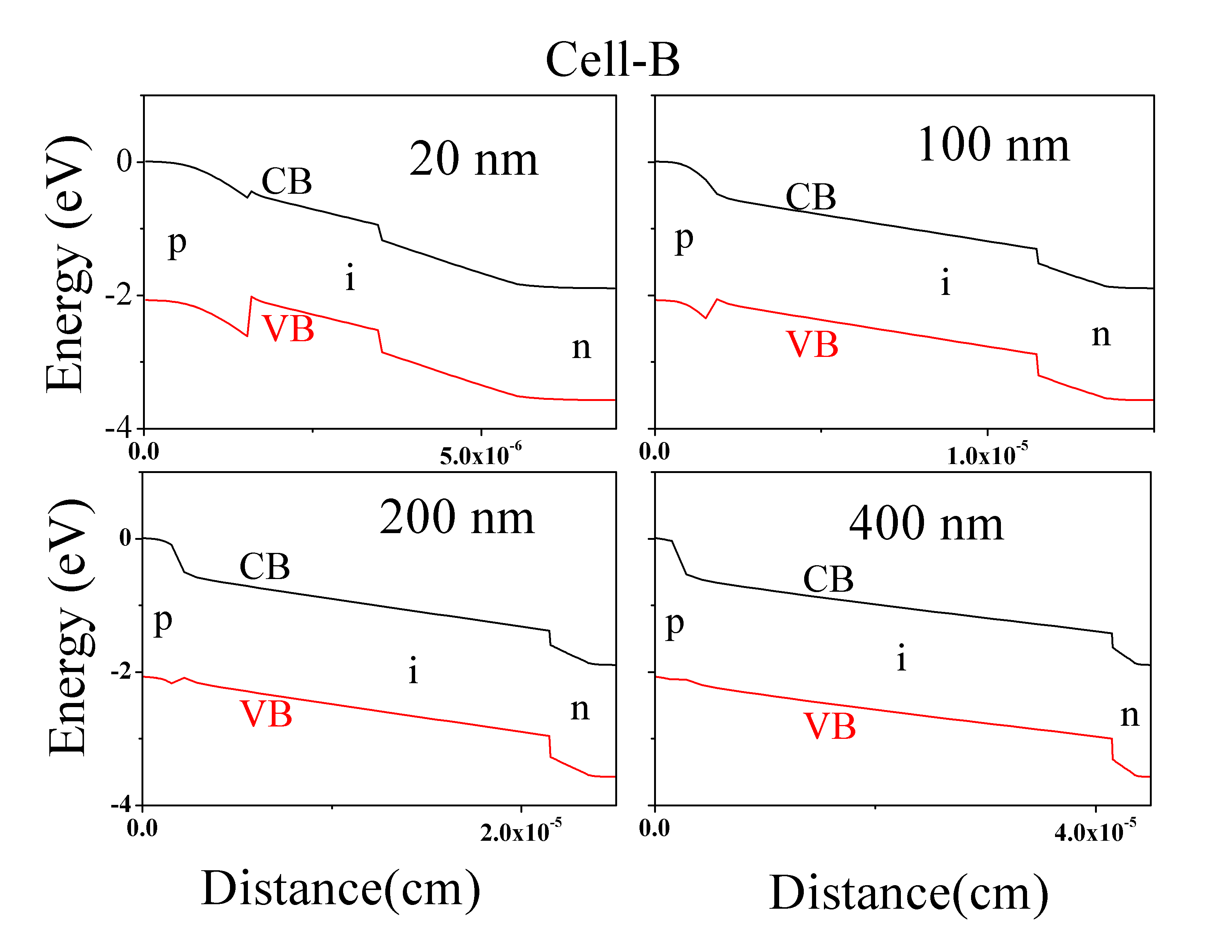}    
	\caption[(a)]{  Energy band diagrams of (a) Cell-A and (b) Cell-B. Here the energy is shown with reference to the p-layer conduction band (CB) mobility edge. The VB indicates valence band mobility edge. }
	\label{fig:fig13}
\end{figure}

\begin{figure}
	\centering
	\includegraphics[width=7cm]{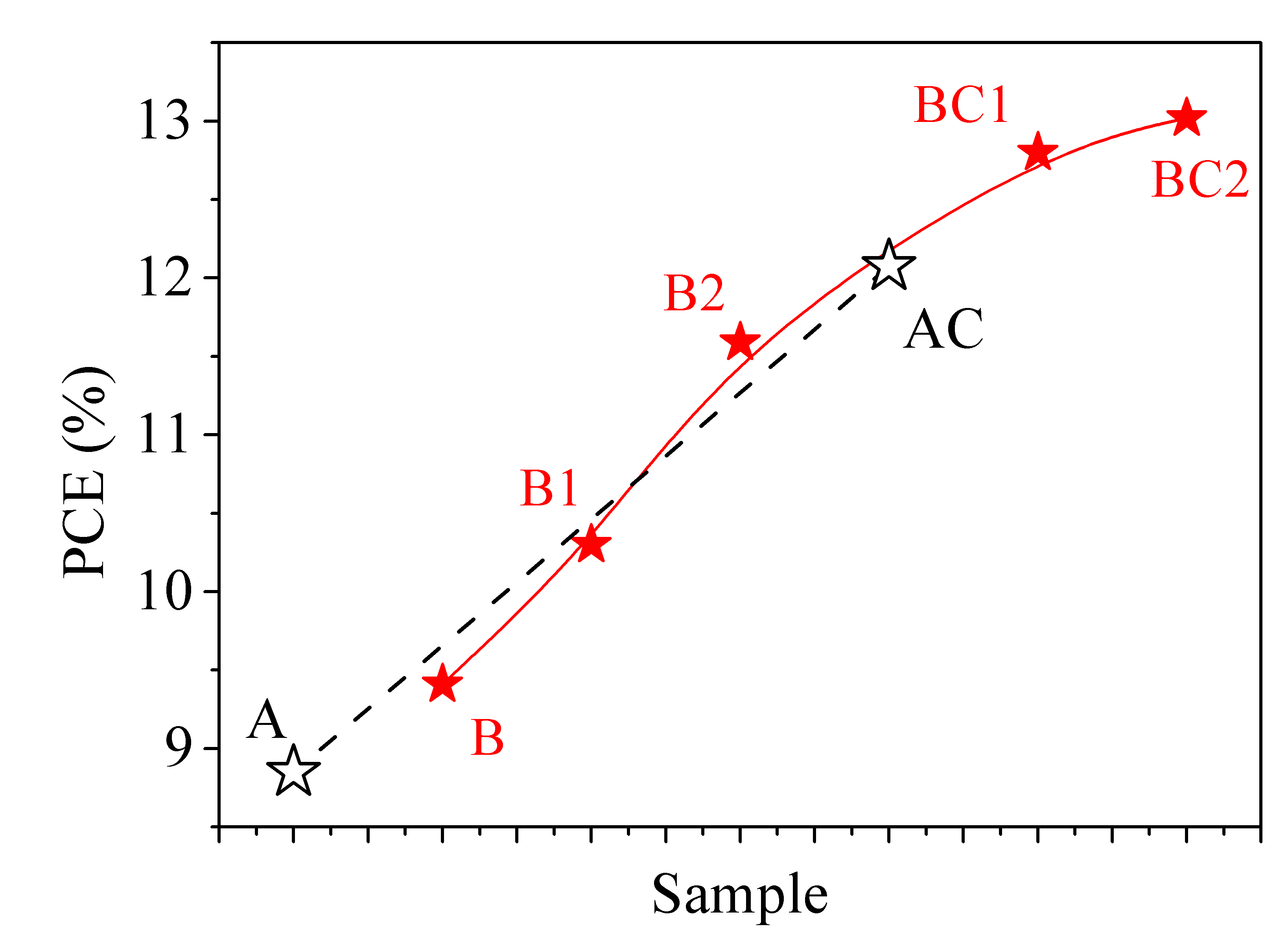} 
	\caption[(a)]{  Trend in variation in PCE of the devices. Improvement in the PCE was estimated from the reference device A to B, when p-nc-SiO:H was removed, but with $d_{i}$ =300 nm. All the subsequent B-type cells were without the p-nc-SiO:H layer. The Cell-B1 was with $d_{i}$ =100 nm, $d_{p}$ = 15nm. The Cell-B2 was with $d_{p}$ = 7nm, with $d_{i}$ =300 nm. The Cell-AC was with the cell similar to the structure of Cell-A, while the active layer defect density here is $6\times 10^{15}$ cm$^{-3}$ . The Cell-BC1 corresponds to the Cell-B1 while the active layer defect density is $6\times 10^{15}$ cm$^{-3}$ . and the Cell-BC2 is similar to that of Cell-BC1, but with 7 nm thick p-a-SiO:H. }
	\label{fig:fig14}
\end{figure}

 The simulated J-V characteristic curves of Cell-AC and Cell-BC are shown in Fig.10. In comparison to the Fig. 6, a significant improvement in characteristic curves is visible here It indicates a similar trend that the $J_{sc}$ increases while the $V_{oc}$ decreases with the increase in the $d_{i} $. The device parameters extracted from these curves are shown in Fig. 11, where such a trend is clearly visible. Here the parameters for the reference device (Cell-A) are shown as a star symbol. It shows that the $V_{oc}$, FF and PCE of the low defect device is higher than the reference device under the same $d_{i} $ = 300nm. The changes are $V_{oc}$ = from 0.836V to 0.953V, $J_{sc}$ = from 15.52mA/cm$^{2}$ to 15.65 mA/cm$^{2}$, FF = from $69.28\%$ to $80.91\%$, PCE = from $8.99\%$ to $12.07\%$. In this condition the best performing device has $d_{i} $ = 300nm. So if the quality of the active layer is improved by reducing its defect density, then a thicker active layer can give higher PCE.

It may be noted that the $V_{oc}$ increases when thickness of the active layer decreases. The reason can be the dependence of the $V_{oc}$ on total number of defects that exists within the active layer. A thicker active layer will have a larger volume than that of a thinner one. As a result, for a thinner active layer solar cell, the photo generated electron hole pairs can be collected at the doped layer more efficiently leading to a higher $V_{oc}$ . For comparison the variation in the $V_{oc}$ due to material defect the V oc for Cell-A and Cell-AC (Fig. 12(a)), Cell-B and Cell-BC (Fig. 12(b)) are plotted together and shown in Fig. 12. It shows that for a reduced defect density, the $V_{oc}$ remains higher. This also indicates that the $V_{oc}$ is inversely related to the total number of defects that the photo generated carriers face on the average, before being collected at the doped layers.

Figure 12 shows the variation in $V_{oc}$ with thickness of i-layer and its defect density. The Cell-A had $p{1}$ , $p{2}$ layers, while the Cell-B had the $p{1}$ -layer only. The results, as shown in Fig. 7, 11, indicate that the $V_{oc}$ does not significantly depend on the $p{2}$ layer. Looking into the figure 12, the variation in $V_{oc}$ can empirically be expressed as

\begin{equation}
	V_{oc} = K_{1} /log(d_{i} )
\end{equation}

Similarly the $J_{sc}$ can be expressed as

\begin{equation}
	J_{sc} = K_{2} /log(d_{i} )
\end{equation}

Here $K_{1} , K_{2}$ are constant terms. It is to be noted that these two equation will fit better for lower defect density or Cell-AC, BC, but will fit less for the Cell-A,B.

As shown in Fig. 6-9, that by eliminating the p-nc-SiO:H layer and reducing the thickness of the a-SiO:H layer from 15 nm to 7 nm, its PCE can be improved significantly. So we simulated the device characteristics with only one p-a-SiO:H layer, similar to that used in Fig. 6-9, except the defect density of the active layer reduced from 6x1016 cm$^{-3}$ to 6x1015 cm$^{-3}$ and using $d_{i} $=300nm, $d_{p}$ = 7nm (Cell-BC2). The resultant device shows $V_{oc}$=0.965V, $J_{sc}$=16.91 mA/cm$^{2}$, FF=$80.03\%$, PCE=$13.05\%$.


\subsubsection{Energy band Diagram}

The energy band diagram is expected to correlate the change in the observed V oc with the device width although there is a difference between these two situations, as the Fig. 13 corresponds to the measurement under dark while the V oc was measured under light, but the similarity is that the external electrical bias in both the situations are absent or open circuit condition. The Cell-A and Cell-B do not exhibit any significant difference in the energy band structure where the J sc differs in Fig. 7 (and Fig. 11), for these two types of cells, which is expected to be due to the difference in available light to the active layers.

However, due to the variation in cell width, a high built-in electric field (which is the ratio between energy difference and device width) helps collecting photo generated carriers. Figure 12 shows that the V oc also inversely related to defect density. Another common feature visible in the band diagram is that for a thinner cell the band discontinuity is more prominent than that with a thicker one. Therefore, with a perfect matching of the energy band among the layers, a further improved device performance is expected.

The Fig. 14 shows the sumary of the improvement in PCE of the device at the different stages.  By removing the 10 nm thick p-nc-SiO:H, the PCE of the Cell-B improved by 0.56 percentage point along with an improved blue response, then by optimizing the active layer leads to an additional 0.89 percentage point enhancement in the PCE (Cell-B1). A further reduction of the p-a-SiO:H layer (to 7 nm) raises the PCE further by 1.29 percentage point along with a further improvement in the blue response of the device (Cell-B2). By reducing the defect density of the active layer of Cell-A from 6x1016 cm$^{-3}$ to 6x1015 cm$^{-3}$ (Cell-AC) leads to the PCE of $12.07\%$. Starting from the Cell-BC1 and by reducing its $d_{p}$ to 7nm further raises of the PCE to $13.02\%$. 

Here, in the AFORS HET simulation the estimated total defect density of the active layer of the real cell is found to be (including the tail state defects) 6x1016 cm$^{-3}$, which is higher than the defect density in intrinsic a-Si:H layer (3.2x1016 cm$^{-3}$) [21]. 

The results indicate that even though a significant increase in optical absorption in the active layer was initiated by removing the p-nc-SiO:H layer and making a thinner p-layer, yet the $J_{sc}$ did not increase significantly (Cell-B). This could be due to several reasons. One of them can be higher midgap defects. Therefore high quality material with lower defect density is always desirable, as can be seen in Cell-AC, BC1, BC2.

The results indicate that the solar cell with a defective material should have a thinner active layer for a better performance [21]. This performance can be estimated by using two separate criteria. One is effective optical absorption, so that it absorbs a certain fraction of incident light and transmits the rests. This criterion will be useful for the top sub-cell of a tandem device as the transmitted light can be used by other sub-cells in a multijunction device structure. Secondly, with a defective active layer a significant part of the photo generated carriers will be lost by recombination. This recombination loss can be reduced by using a thinner active layer, although a very thin active layer is not practical.

Increasing Ge content within the film may decrease optical gap and hence increase optical absorption in the longer wavelength region but it simultaneously degrades short wavelength response [22], it may be because of increased carrier loss [11] due to  recombination at the dangling bond defect sites [23]. On the other hand by adapting the approach of modifying device structure, as described here, both the short as well as long wavelength responses can be improved without compromising material properties.

Although this work is based on numerical simulation of solar cell, yet it clearly demonstrates the importance of having a thinner doped window layer with which short as well as long wavelength response of the device can be improved. Generally it is believed that wide band gap doped window layer can
improve the short wavelength response, whereas here the results show that a similar improvement can be achieved by using a thinner window layer. Another important aspect of this investigation is the improvement in long wavelength response of the device by using a thicker active layer. In this respect the dominating belief is that a low band gap active layer is needed to achieve this. Further investigation also indicates that a thicker active layer can give a high PCE if its defect density remains low. Therefore, a
particular design of a solar cell is expected to be dependent on the defect density of the active layer.

\section{Conclusions}
In conclusion, the present analysis shows that optical absorption by the active layer is an important criterion for improving device performance. It also shows that the short and long wavelength response can also be tuned by adapting a suitable device structure; thinner doped window layer is always preferable for reducing parasitic optical absorption and improving short wavelength response. Thicker active layer can be used to raise the long wavelength response but it should be limited to the maximum power conversion efficiency.

\end{document}